\documentclass[prd,aps,twocolumn,superscriptaddress,dvips,10pt]{revtex4}
\usepackage{feynmp}
\usepackage{amsmath}
\usepackage{amssymb}
\usepackage{epsfig}
\usepackage{graphicx}
\usepackage{subfigure}
\usepackage{hyperref}
\usepackage{bm}
\usepackage{color}
\usepackage{lineno}
\usepackage{threeparttable}
\usepackage{supertabular}

\begin{document}

\title{Nonperturbative study of quantum many-body correlation effects in
neutron stars: Equation of state}

\author{Hao-Fu Zhu}
\affiliation{CAS Key Laboratory for Research in Galaxies and
Cosmology, Department of Astronomy, University of Science and
Technology of China, Hefei, Anhui 230026, China}\affiliation{School
of Astronomy and Space Science, University of Science and Technology
of China, Hefei, Anhui 230026, China}
\author{Xufen Wu}
\affiliation{CAS Key Laboratory for Research in Galaxies and
Cosmology, Department of Astronomy, University of Science and
Technology of China, Hefei, Anhui 230026, China}\affiliation{School
of Astronomy and Space Science, University of Science and Technology
of China, Hefei, Anhui 230026, China}
\author{Guo-Zhu Liu}
\altaffiliation{Corresponding author: gzliu@ustc.edu.cn}
\affiliation{Department of Modern Physics, University of Science and
Technology of China, Hefei, Anhui 230026, P. R. China}

\begin{abstract}
Although neutron stars have been studied for decades, their internal
structure remains enigmatic, mainly due to large uncertainties in
the equation of state. In neutron stars, the nucleons are strongly
interacting by exchanging mesons, which can lead to significant
quantum many-body correlation effects. Mean-field calculations
failed to capture these effects. Here, we develop a nonperturbative
quantum field-theoretic approach to handle strongly correlated dense
nuclear matter within the framework of quantum hadrodynamics. We
show that the many-body effects can be incorporated in the
Dyson-Schwinger equation of the nucleon propagator. Based on a
linear $\sigma$-$\omega$-$\rho$ model, we successfully reproduce six
empirical observable quantities of saturation nuclear matter by
tuning six parameters. After including the many-body effects into
the equation of state of realistic neutron star matter, we obtain a
mass-radius relation that is comparable with recent astrophysical
observations of neutron stars.
\end{abstract}

\maketitle

%%%%%%%%%%%%%%%%%%%%%%%%%%%%%Main Body%%%%%%%%%%%%%%%%%%%%%%%%%%%%%%%%%%%%%

\section{Introduction}\label{sec:introduction}

The physical properties of neutron stars are characterized by a
series of observable quantities \cite{Glendenningbook, Lattimer04},
including the maximal mass, mass-radius ($M$-$R$) relation, tidal
deformability, moment of inertia, gravitational shift, and so on.
Such quantities depend crucially on the equation of state (EOS),
which relates the energy density $\epsilon$ to the pressure $P$
\cite{Lattimer16, Burgio21}. Unfortunately, the neutron star EOS has
not been determined. Although the internal composition is still in
debate, the most common view \cite{Glendenningbook, Lattimer04} is
that the neutron star core is dominantly composed of neutrons
strongly interacting with each other by exchanging various mesons.
To ensure causality, it is convenient to use relativistic quantum
field theories to describe nucleon-meson (NM) interactions, which
defines the framework of quantum hadrodynamics (QHD)
\cite{Walecka74, Glendenningbook, Boguta77, Serot79, Kubis97,
Reinhard89, Dutra14, Dutra16, NL3, GM1, TM1, NL3omegarho, FSUGold,
BigApple, TW99, DDME2, DD2, DDVT, DDMEX}.

According to the extensive investigations of quantum many-particle
systems \cite{Waleckabook}, one can affirm that strong NM
interactions induce significant quantum correlation effects, such as
the Landau damping of nucleon states, the modification of nucleon
velocity, and the renormalization of nucleon masses. These effects
could markedly change the bulk properties of nuclear matter, and
thus should be properly included in the EOS. Unfortunately, such
effects are poorly considered in previous works on QHD models. Since
the NM interactions are quite strong, ordinary weak perturbation
theory is failed. In practice, relativistic mean-field theory (RMFT)
has been widely used to study QHD models \cite{Glendenningbook,
Walecka74, Boguta77, Reinhard89, Dutra14, Dutra16, Serot79, Kubis97,
NL3, GM1, TM1, NL3omegarho, FSUGold, TW99, BigApple, DDME2, DD2,
DDVT, DDMEX}. RMFT has achieved remarkable success in the
theoretical description of finite nuclei. However, the nucleon
density in the core region of neutron star is much higher than in
finite nuclei. As a result, the quantum fluctuation and the
correlation effects are substantially enhanced. RMFT, as a
semi-classical method, cannot incorporate such quantum fluctuation
and correlation effects. Furthermore, RMFT relies on intricate
interactions with many tuning parameters, and often suffers from the
instability difficulty \cite{Glendenningbook, Reinhard89}.

Apart from RMFT, chiral effective field theory ($\chi$EFT)
\cite{Meissner, Machleidt} and the quantum Monte Carlo (QMC) method
\cite{Carlson15} are also frequently applied to study nuclear
matter. These two methods can be combined to gain useful
constraints on the neutron star EOS \cite{Tews18, Huth22}.
Nevertheless, $\chi$EFT results are reliable only at densities lower
than $2\rho^{}_{\mathrm{B}0}$, where $\rho^{}_{\mathrm{B}0}$ is the
nuclear saturation density. Hence, the combination of $\chi$EFT and
QMC simulation becomes questionable at high densities \cite{Tews18,
Huth22}. The QMC method also has its limitations. It usually uses a
small number of particles in a box to mimic infinite nuclear
matter. The validity of this scheme needs to be justified. The QMC
simulations also suffer from the fermion sign problem and finite-size
effects at high densities.

On the experimental side, astrophysical observations have made
significant progress in the past decade. The detection of
gravitational waves by the LIGO Scientific and Virgo Collaborations,
especially the GW170817 event from the merger of binary neutron
stars, has led us to a new era of multimessenger astronomy
\cite{Abbott17, Abbott18}. The Neutron Star Interior Composition
Explorer (NICER) has enabled the simultaneous measurement of the
masses and radii of some neutron stars \cite{Miller19, Riley19,
Miller21, Riley21}. The discovery of several massive neutron stars
with masses $> 2.0M_{\odot}$, where $M_{\odot}$ is the solar mass,
raised the critical question as to what is the upper limit of
neutron star mass \cite{Demorest10, Antoniadis13, Cromartie19,
Romani22}. These exciting findings provide exceptional opportunities
to examine the effects of many-body correlations on the EOS of
neutron stars.

In this paper, we propose a nonperturbative quantum field-theoretic
approach to handle the strong coupling regime of QHD. This approach
surmounts the flaws of weak perturbation and mean-field theories,
and is valid within a wide range of nucleon density. We shall show
that the effects of many-body correlations can be amply incorporated
in the Dyson-Schwinger (DS) equation \cite{Itzykson, Roberts00,
Liu21, Pan21} of the renormalized nucleon propagator $G(k)$, where
$k\equiv(\varepsilon,\mathbf{k})$ is the four-momentum. After
performing a nonperturbative analysis of a linear
$\sigma$-$\omega$-$\rho$ model that contains six tuning parameters,
we acquire a perfect fit to the experimental results of six
observable quantities of saturation nuclear matter, including the
saturation density $\rho^{}_{\mathrm{B}0}$, binding energy per
nucleon $E_{\mathrm{b}}$, effective nucleon mass
$m^{\ast}_{\mathrm{N}}$, compressible modulus $K$, symmetry energy
$E_{\mathrm{s}}$, and symmetry energy slope $L_{\mathrm{s}}$.

The simulated model parameters are then adopted to compute the
neutron star EOS. In reality, pure neutron matter is not stable
\cite{Glendenningbook}. A small fraction of protons, electrons, and
muons should be present in neutron star cores. They coexist with
neutrons to sustain the $\beta$ equilibrium as well as the
electrical neutrality \cite{Glendenningbook}. The quantum many-body
effects caused by the proton-meson interactions can be determined by
using our approach. The resulting total EOS of realistic neutron
star matter leads to an $M$-$R$ relation that is dependent on the
value of $L_{\mathrm{s}}$. When $L_{\mathrm{s}}$ is tuned to be
roughly $\approx (30-50)~\mathrm{MeV}$, the radii of low-mass neutron
stars produced by our EOS is comparable with astrophysical
observations. The corresponding maximal mass of neutron stars is
about 2.67$~M_{\odot}$, which, however, might be modified if
additional degrees of freedom are considered.

The rest of the paper is organized as follows. In
Sec.~\ref{sec:dse}, we give the DS equation of the renormalized
neutron propagator and decompose it into three self-consistent
integral equations. In Sec.~\ref{sec:fitting}, we derive the
expressions of some observable quantities of saturation nuclear
matter and compare the results with experiments. In
Sec.~\ref{sec:eosofnss}, we obtain the EOS that includes the impact
of nonperturbative effects. The results are summarized in
Sec.~\ref{sec:summary} with a highlight of future research work.

\section{Dyson-Schwinger equation of neutron propagator}\label{sec:dse}

As the first step, we should define an appropriate QHD model as the
starting point. The simplest QHD model is the linear
$\sigma$-$\omega$ model, also known as linear Walecka model
\cite{Walecka74}. Given that perturbation theory is invalid, Walecka
employed RMFT to address the linear $\sigma$-$\omega$ model and
reproduced the saturation density $\rho_{B0}$ and the binding energy
per nucleon $E_{b}$ by tuning two Yukawa coupling parameters
$g_{\sigma}$ and $g_{\omega}$. However, the RMFT results of other
nuclear-matter quantities were at odds with experiments
\cite{Glendenningbook}. To remedy this flaw, Bodmer and Bogutta
\cite{Boguta77} included nonlinear $-b_{0}\sigma^{3}$ and
$-c_{0}\sigma^{4}$ terms to linear $\sigma$-$\omega$ model.
Afterwards, more mesons, such as $\rho$ \cite{Serot79} and $\delta$
\cite{Kubis97}, along with many more free parameters, were
introduced to gain a better fitting to the experimental data of
nuclear quantities and neutron star observations.

In the past several decades, RMFT has been applied to study not only
finite nuclei but also infinite nuclear matter and neutron stars.
Despite such success, RMFT and its variants have obvious limitations
\cite{Reinhard89}. First of all, RMFT is essentially semiclassical
as it ignores all the quantum fluctuations of meson fields. In
addition, almost all RMFT models contain two nonlinear self-coupling
terms of $\sigma$ mesons: $-b_{0}\sigma^3$ and $-c_{0}\sigma^4$.
Such terms make a contribution $U(\sigma) = b_{0}\sigma^3 +
c_{0}\sigma^4$ to the total energy of the system. In many widely
used RMFT models, $b_{0}$ and/or $c_{0}$ take negative values. In
that case, the total energy has no lower bound, implying that the
systems described by such RMFT models are unstable and would undergo
thermodynamic instability once quantum fluctuations become strong
enough at high densities. This inconsistency cast serious doubt on
the validity of RMFT results. Another problem is that RMFT often
needs to manually introduce some intricate interactions. The
resultant RMFT models are very complicated, making it difficult to
carry out systematic theoretic calculations. These limitations might
not be a big problem in finite nuclei where the nucleon density is
relatively low. However, quantum fluctuations are greatly enhanced
and the nucleon-meson interactions become quite strong in the
high-density cores of neutron stars. The quantum many-body effects
should be incorporated into the EOS.

We consider a generic $\sigma$-$\omega$-$\rho$-$\delta$ QHD model
\cite{Walecka74, Glendenningbook}
\begin{eqnarray}
\mathcal{L} &=& \overline{\psi}\left(i
\partial_{\mu}\gamma^{\mu} - m^{}_{\mathrm{N}}\right)\psi + \frac{1}{2}
\partial_{\mu}\sigma\partial^{\mu}\sigma - \frac{1}{2}m^{2}_{\sigma}
\sigma^{2} \nonumber \\
&& -\frac{1}{4}\omega_{\mu\nu}\omega^{\mu\nu} +
\frac{1}{2} m^{2}_{\omega}\omega_{\mu}\omega^{\mu} \nonumber \\
&& -\frac{1}{4}\bm{\rho}_{\mu\nu}\cdot\bm{\rho}^{\mu\nu} +
\frac{1}{2}m^{2}_{\rho}\bm{\rho}_{\mu}\cdot\bm{\rho}^{\mu}\nonumber \\
&& +\frac{1}{2}\partial_{\mu}\bm{\delta}\cdot \partial^{\mu}
\bm{\delta} - \frac{1}{2}m^{2}_{\delta}\bm{\delta}^{2} \nonumber \\
&& +g_{\sigma}\sigma\bar{\psi}\psi - g_{\omega}\omega_\mu
\overline{\psi}\gamma^{\mu}\psi \nonumber \\
&&-\frac{1}{2}g_{\rho}\bm{\rho}_{\mu}\cdot \bar{\psi}\bm{\tau}
\gamma^{\mu}\psi+ g_{\delta}\bm{\delta}\cdot \bar{\psi}
\bm{\tau}\psi, \label{eq:LWmodel}
\end{eqnarray}
where $\partial_{\mu}=(\partial_{t},\bm{\partial})$. The tensors
$\omega_{\mu\nu}$ and $\bm{\rho}_{\mu\nu}$ are
\begin{eqnarray}
\omega_{\mu\nu} &=& \partial_\mu\omega_{\nu} -
\partial_{\nu}\omega_\mu, \\
\bm{\rho}_{\mu\nu} &=& \partial_\mu \bm{\rho}_{\nu} -
\partial_{\nu}\bm{\rho}_\mu-g_{\rho}\bm{\rho}_\mu\times
\bm{\rho}_{\nu}.
\end{eqnarray}
The spinor $\psi$, whose conjugate is $\bar{\psi} = \psi^{\dag}
\gamma^{0}$, has eight components for nuclear matter, namely $\psi
= (\psi_\mathrm{p}, \psi_\mathrm{n})^{T}$. For a pure neutron
matter, $\psi=\psi_n$ has only four components. Nucleons couple to
neutral $\sigma$ mesons, denoted by an isoscalar scalar field
$\sigma$, neutral vector $\omega$ mesons, denoted by an isoscalar
vector field $\omega^{\mu}$, charged vector $\rho$ mesons, denoted
by an isovector vector field $\bm{\rho}^{\mu}$, and charged
$\delta$ mesons, denoted by an isovector scalar field $\bm{\delta}$.
The Pauli matrices $\bm{\tau}$ operate in the isospin space.
Considering the rotational invariance around the third axis in
isospin space, we only retain the isospin three-component of
$\rho_{3}^{\mu}$ and that of $\delta_{3}$, namely neutral $\rho_{0}$
and neutral $\delta_{0}$. The bare nucleon mass is $m_{\mathrm{N}} =
939~$MeV and the rest meson masses are $m_{\sigma} = 550~$MeV,
$m_{\omega} = 783~$MeV, $m_{\rho} = 763~$MeV, and $m_{\delta}=980~$MeV.
Notice that this model does not contain self-couplings nor
cross-couplings of mesons. This avoids the instability difficulty.

At the mean-field level, one replaces the quantum fields $\sigma$,
$\omega$, $\rho$, and $\delta$ with their mean values $\langle \sigma
\rangle$, $\langle \omega \rangle$, $\langle \rho \rangle$, and
$\langle \delta \rangle$, respectively. Then the above QHD model can
be treated using the standard RMFT procedure \cite{Glendenningbook}.
As mentioned at the beginning of this section, such an approximation
is not satisfactory. To illustrate the importance of surpassing
mean-field theory and developing nonperturbative methods, it is
instructive to consider the historical development of the
microscopic theory of superconductivity as a reference. While
Bardeen-Cooper-Schrieffer (BCS) theory \cite{BCS} can qualitatively
explain the zero resistivity and Meissner effect of
superconductivity, the observable quantities, especially the
critical temperature $T_{c}$, computed based on BCS mean-field
theory are unreliable for most materials. To compute observable
quantities more precisely, it is necessary to consider some major
quantum many-body effects, such as the retardation of phonon
propagation and the electron damping, ignored by BCS theory. These
effects can be efficiently handled by adopting the nonperturbative
Eliashberg theory \cite{MEtheory}, which treats the electron-phonon
interaction in a self-consistent manner and computes $T_{c}$ by
solving a number of nonlinear integral equations. Over the last
60 years, the Eliashberg theory has been the standard microscopic
theory of phonon-mediated superconductivity. For many
superconductors, the results of $T_{c}$ obtained by Eliashberg
theory are in good agreement with experiments.

Inspired by the great success of Eliashberg theory as a
well-justified substitute of BCS mean-field theory, here we employ a
nonperturbative approach to handle the strongly correlated dense
nuclear matter. Different from RMFT, we view the model of
Eq.~(\ref{eq:LWmodel}) as a fully fledged quantum field theory and
treat the nucleon dynamics and meson dynamics on an equal footing.
We find that the nonperturbative effects of QHD models can be taken
into account by solving the self-consistent DS equation of nucleon
propagator $G(k)$. This sort of approach has been exploited to study
dynamical chiral symmetry breaking in particle physics
\cite{Roberts00} and superconductivity \cite{Liu21} and excitonic
insulating transition \cite{Pan21} in condensed matter physics. As
far we know, it has not been adopted to compute the EOS and $M$-$R$
relation of neutron stars.

\begin{figure}[htbp]
\centering
\includegraphics[width=3.5in]{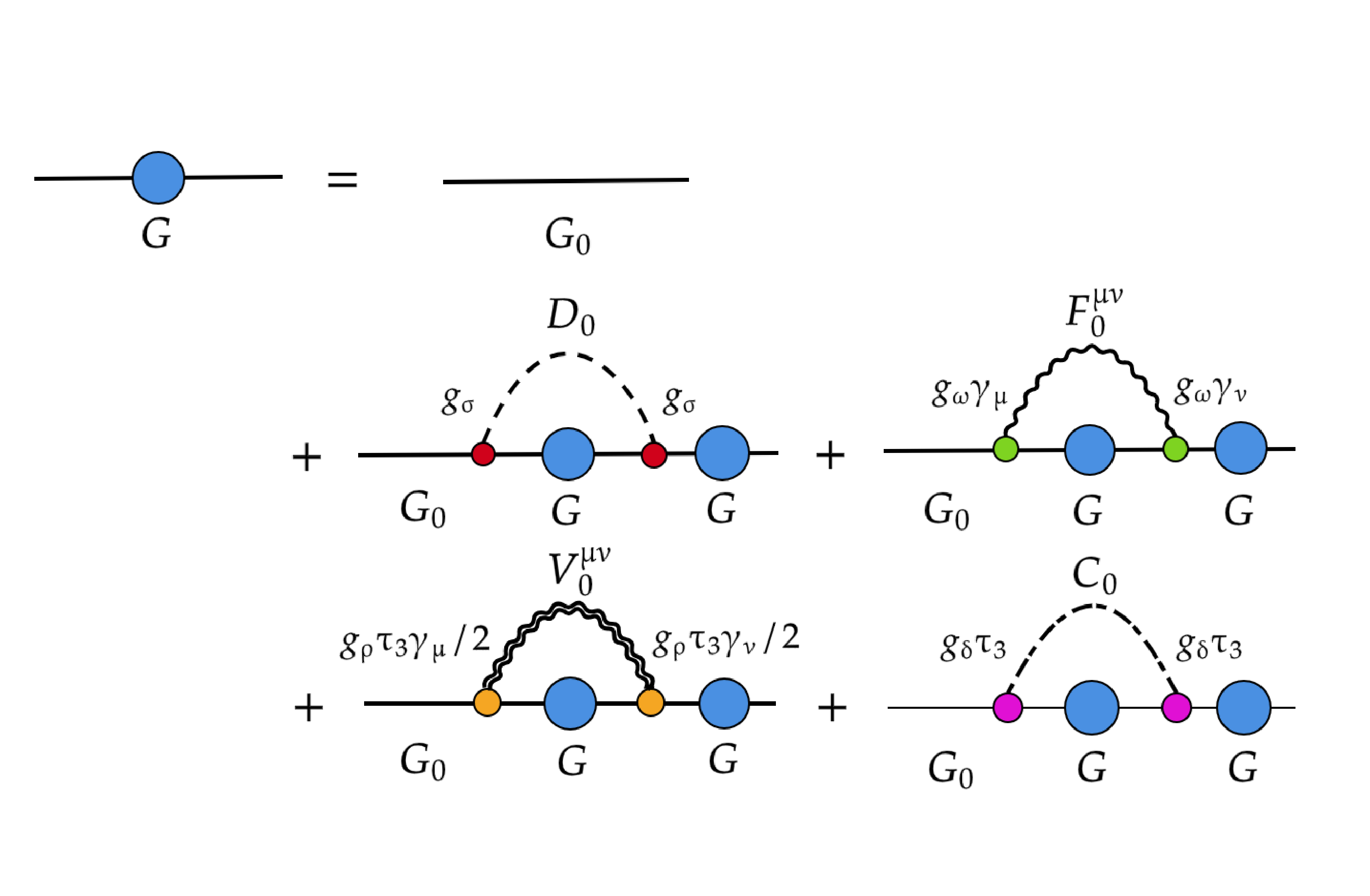}
\caption{Diagrammatic illustration of the DS equation of $G(k)$.}
\label{fig:Feymann}
\end{figure}

According to the analysis of the Appendix \ref{sec:dsederivation}, the
DS equation of renormalized nucleon propagator $G(k)$ is
\begin{eqnarray}
G^{-1}(k) &=& G_{0}^{-1}(k) + ig_{\sigma}^{2}\int
\frac{d^4q}{(2\pi)^4}G(k+q)D_0(q) \nonumber \\
&& - ig_{\omega}^{2}\int\frac{d^4q}{(2\pi)^4}\gamma_{\mu}
G(k+q)F_0^{\mu\nu}(q)\gamma_{\nu}\nonumber \\
&& -i\frac{g^{2}_{\rho}}{4}\int \frac{d^4q}{(2\pi)^4}\tau_{3}
\gamma_{\mu}G(k+q) V^{\mu\nu}_{0}(q)\tau_{3}\gamma_{\nu}
\nonumber \\
&& -ig^{2}_{\delta}\int \frac{d^4q}{(2\pi)^4}\tau_{3}
G(k+q)C_{0}(q)\tau_{3}. \label{eq:dselwmodel}
\end{eqnarray}
A schematic Feynman diagram of this equation is shown in
Fig.~\ref{fig:Feymann}. This DS equation contains, apart from
$G(k)$, the free neutron propagator
\begin{eqnarray}
G_{0}(k) = \frac{1}{k_{\mu}\gamma^{\mu}-m^{}_{\mathrm{N}}},
\end{eqnarray}
and three free propagators of $\sigma$, $\omega$, $\rho_0$, and
$\delta_0$ mesons:
\begin{eqnarray}
D_0(q) &=& \frac{1}{q^{2}-m^{2}_{\sigma}}, \\
F^{\mu\nu}_0(q) &=& -\frac{1}{{q}^{2}-m^{2}_{\omega}}
\left(g^{\mu\nu} -
\frac{{q}^{\mu}{q}^{\nu}}{m^{2}_{\omega}}\right), \\
V^{\mu\nu}_0(q) &=& -\frac{1}{{q}^{2}-m^{2}_{\rho}}\left(
g^{\mu\nu}-\frac{{q}^{\mu}{q}^{\nu}}{m^{2}_{\rho}}\right), \\
C_0(q) &=& \frac{1}{q^{2}-m^{2}_{\delta}}.
\end{eqnarray}
The four-momenta of neutrons and mesons are
$k\equiv(\varepsilon,\mathbf{k})$ and $q\equiv(\omega,\mathbf{q})$,
respectively. The meson propagators depend on both energy and
momentum, so the dynamics of all mesons is incorporated in the DS
equation. The values of coupling parameters $g_{\sigma}$,
$g_{\omega}$, $g_{\rho}$, and $g_{\delta}$ should be determined by
comparing theoretical results with the experimental data of the
saturation nuclear matter.

The Fermi energy $E^{~}_{\mathrm{F}}$ provides a natural energy
scale and can be used to define the integration range of $\omega$.
Here, we choose $\omega \in
[-\Omega_{\mathrm{c}},+\Omega_{\mathrm{c}}]$, where
$\Omega_{\mathrm{c}}=1000~\mathrm{MeV}$ is of the same order of
$E^{~}_{\mathrm{F}}$ at $6\rho^{~}_{\mathrm{B}0}$. The absolute
value of meson momentum $|\mathbf{q}|$ lies within the range of
$[0,\Lambda_{\mathrm{c}}k^{~}_{\mathrm{F}}]$, where
$k^{~}_{\mathrm{F}}$ is the Fermi momentum and
$\Lambda_{\mathrm{c}}$ is a positive tuning parameter. Thus, the
integral measure is
\begin{eqnarray}
\int \frac{d^4q}{(2\pi)^4} \equiv
\int_{-\Omega_{\mathrm{c}}}^{+\Omega_{\mathrm{c}}}
\frac{d\omega}{2\pi}\int_{0}^{\Lambda_{\mathrm{c}}
k^{~}_{\mathrm{F}}} \frac{d^{3}\mathbf{q}}{(2\pi)^{3}}.
\end{eqnarray}
All the results are free of divergences, thus it is not necessary to
perform renormalization calculations.

As mentioned, our model does not contain self- and
cross-coupling terms of the meson fields. However, we still wish to
include their potential contributions. Generic field-theoretic
analysis shows that such coupling terms can renormalize meson
masses. In light of this, we replace the bare mass $m_{\sigma}$
appearing in Eq.~(\ref{eq:LWmodel}) and $D_0(q)$ with the
renormalized mass $m^{\ast}_{\sigma}$ and take the ratio
$m^{\ast}_{\sigma}/m_{\sigma}$ as an independent tuning parameter.
One may let $m_{\omega}$ vary, but tuning $m_{\sigma}$ is
sufficient.

On generic grounds, one can expand $G(k)$ as
\begin{eqnarray}
G(k) = \frac{1}{A_0(k) \varepsilon \gamma^0 - A_1(k)\mathbf{k}
\cdot\bm{\gamma} - A_2(k)m^{~}_{\mathrm{N}}},
\label{eq:genericformgp}
\end{eqnarray}
where $A_{0}(k)$, $A_{1}(k)$, and $A_{2}(k)$ are the renormalization
functions of the wave function (field operator), the neutron
velocity, and the neutron mass, respectively. The effects of quantum
many-body correlation are embodied in these three functions. In the
non-interacting limit, $A_{0}(k)=A_{1}(k)=A_{2}(k) = 1$. The
many-body effects are characterized by the deviation of
$A_{0,1,2}(k)$ from unity.

The DS equation is formally very complicated. To simplify
calculations, we make some further approximations. On account of the
translational invariance and rotational symmetry of infinite nuclear
matter, we retain only the contribution of the time component of
$\omega^\mu$ and $\rho^\mu_3$, which is achieved by taking
\begin{eqnarray}
&&\gamma_{\mu}\rightarrow \gamma_0, \quad
\tau_3\gamma_{\mu}\rightarrow \tau_3\gamma_0, \\
&&F_0^{\mu\nu}(q)\approx F_{0}^{00}(q), \quad
V_{0}^{\mu\nu}(q)\approx V_0^{00}(q).
\end{eqnarray}
Moreover, we drop the terms $\frac{q^{\mu}q^{\nu}}{m^{2}_{\omega}}$
from $F_{0}^{\mu\nu}(q)$ and $\frac{q^{\mu}q^{\nu}}{m^{2}_{\rho_0}}$
from $V_{0}^{\mu\nu}(q)$ to preserve the baryon number conservation
and isospin conservation, respectively. Under such approximations,
we substitute the generic propagator (\ref{eq:genericformgp})
into the DS equation (\ref{eq:dselwmodel}), and then find
\begin{widetext}
\begin{eqnarray}
&&A_0(k)\varepsilon\gamma^0 - A_1(k)\mathbf{k}\cdot \bm{\gamma} -
A_2(k)m^{~}_{\mathrm{N}} -
\varepsilon\gamma^0+\mathbf{k}\cdot\bm{\gamma} +
m_{\mathrm{N}} \nonumber \\
&=& -ig^{2}_{\sigma} \int \frac{d^4q}{(2\pi)^4} \frac{1}{A_0(k+q)
(\varepsilon+\omega)\gamma^{0} - A_1(k+q)\mathbf{(k+q)}
\cdot\bm{\gamma}-A_2(k+q)m^{~}_{\mathrm{N}}}D_0(q) \nonumber \\
&&-ig^{2}_{\omega}\int\frac{d^4q}{(2\pi)^4}\gamma_0
\frac{1}{A_0(k+q) (\varepsilon+\omega) \gamma^0-
A_1(k+q)\mathbf{(k+q)}\cdot \bm{\gamma} -
A_2(k+q)m^{~}_{\mathrm{N}}}F_0^{00}(q)\gamma_0 \nonumber \\
&&-i\frac{g^{2}_{\rho}}{4} \int \frac{d^4q}{(2\pi)^4}
\tau_3\gamma_0\frac{1}{A_0(k+q)(\varepsilon+\omega)\gamma^0-
A_1(k+q)\mathbf{(k+q)}\cdot\bm{\gamma} -
A_2(k+q)m_{\mathrm{N}}}V_0^{00}(q)\tau_3\gamma_0 \nonumber \\
&&-ig^{2}_{\delta} \int \frac{d^4q}{(2\pi)^4}\tau_3
\frac{1}{A_0(k+q) (\varepsilon+\omega)\gamma^0 -
A_1(k+q)\mathbf{(k+q)} \cdot\bm{\gamma} -
A_2(k+q)m^{~}_{\mathrm{N}}}C_0(q)\tau_3. \label{eq:integraleq}
%\nonumber \\
\end{eqnarray}

This single equation can be decomposed into three integral
equations. For instance, multiplying $\gamma^0$ and $\bm{\gamma}$ to
both sides of Eq.~(\ref{eq:integraleq}) and then calculating the
trace, one would get the equations for $A_0(k)$ and $A_1(k)$,
respectively. Computing the trace of both sides of
Eq.~(\ref{eq:integraleq}) yields the equation of $A_2(k)$. These
three equations are nonlinear and coupled to each other. Solving
them requires enormous computational resources. The computational
time can be substantially reduced if these equations have only one
integration variable. One could ignore either the momentum
dependence or the energy dependence of $A_0(k)$, $A_1(k)$, and
$A_2(k)$. According to our numerical calculations, $A_0(k)$,
$A_1(k)$, and $A_2(k)$ exhibit a rather weak dependence on the
momentum $\mathbf{k}$ for a fixed energy. It is therefore justified
to fix their $\mathbf{k}$ at Fermi momentum
$\mathbf{k}^{~}_{\mathrm{F}}$ and set $A_{0,1,2}(\varepsilon,
\mathbf{k}^{~}_{\mathrm{F}}) \equiv A_{0,1,2}(\varepsilon)$. We
emphasize here that this approximation does not entirely neglect the
momentum dependence. In particular, the Fermi surface is still
present and Pauli's exclusion principle continues to play an
essential role. The nonlinear integral equations of
$A_0(\varepsilon)$, $A_1(\varepsilon)$, and $A_2(\varepsilon)$ are
\begin{eqnarray}
A_0(\varepsilon) &=& 1-\frac{i}{\varepsilon}\int\frac{d\omega
d^3\mathbf{q}}{(2\pi)^4}\frac{A_0(\varepsilon+\omega)(\varepsilon
+\omega)}{A^{2}_0(\varepsilon+\omega)(\varepsilon+\omega)^{2} -
A^{2}_1(\varepsilon+\omega)(\mathbf{k}+\mathbf{q})^{2} -
A^{2}_2(\varepsilon+\omega)m^{2}_\mathrm{N}} \nonumber \\
&&\times\Bigg(\frac{g^{2}_{\sigma}}{\omega^{2}-\mathbf{q}^{2}-m^{2}_{\sigma}}
- \frac{g^{2}_{\omega}}{\omega^{2}-\mathbf{q}^{2}-m^{2}_{\omega}} -
\frac{g^{2}_{\rho}/4 }{\omega^{2}-\mathbf{q}^{2}-m^{2}_{\rho}} +
\frac{g^{2}_{\delta}}{\omega^{2}-\mathbf{q}^{2}-m^{2}_{\delta}}\Bigg),
\label{eq:integraleqA0} \\
A_1(\varepsilon)&=& 1-\frac{i}{|\mathbf{k}|}\int\frac{d\omega
d^3\mathbf{q}}{(2\pi)^4}\frac{A_1(\varepsilon+\omega)|\mathbf{k} +
\mathbf{q}|}{A^{2}_0(\varepsilon+\omega)(\varepsilon+\omega)^{2} -
A^{2}_1(\varepsilon+\omega)(\mathbf{k}+\mathbf{q})^{2} -
A^{2}_2(\varepsilon+\omega)m^{2}_\mathrm{N}} \nonumber \\
&&\times\Bigg(\frac{g^{2}_{\sigma}}{\omega^{2}-\mathbf{q}^{2} -
m^{2}_{\sigma}} +
\frac{g^{2}_{\omega}}{\omega^{2}-\mathbf{q}^{2}-m^{2}_{\omega}} +
\frac{g^{2}_{\rho}/4}{\omega^{2}-\mathbf{q}^{2}-m^{2}_{\rho}} +
\frac{g^{2}_{\delta}}{\omega^{2}-\mathbf{q}^{2}-m^{2}_{\delta}}\Bigg),
\label{eq:integraleqA1}\\
A_2(\varepsilon)&=& 1+\frac{i}{m_{\mathrm{N}}}\int \frac{d\omega
d^3\mathbf{q}}{(2\pi)^4}\frac{A_2(\varepsilon+\omega)
m_{\mathrm{N}}}{A^{2}_0(\varepsilon+\omega)(\varepsilon+\omega)^{2}
- A^{2}_1(\varepsilon+\omega)(\mathbf{k}+\mathbf{q})^{2} -
A^{2}_2(\varepsilon+\omega)m^{2}_\mathrm{N}} \nonumber \\
&&\times\Bigg(\frac{g^{2}_{\sigma}}{\omega^{2}-\mathbf{q}^{2} -
m^{2}_{\sigma}} - \frac{g^{2}_{\omega}}{\omega^{2} -
\mathbf{q}^{2}-m^{2}_{\omega}} - \frac{g^{2}_{\rho}/4}{\omega^{2} -
\mathbf{q}^{2}-m^{2}_{\rho}}+\frac{g^{2}_{\delta}}{\omega^{2} -
\mathbf{q}^{2}-m^{2}_{\delta}}\Bigg).\label{eq:integraleqA2}
\end{eqnarray}
\end{widetext}

\begin{figure}[htbp]
\centering
\includegraphics[width=3.98in]{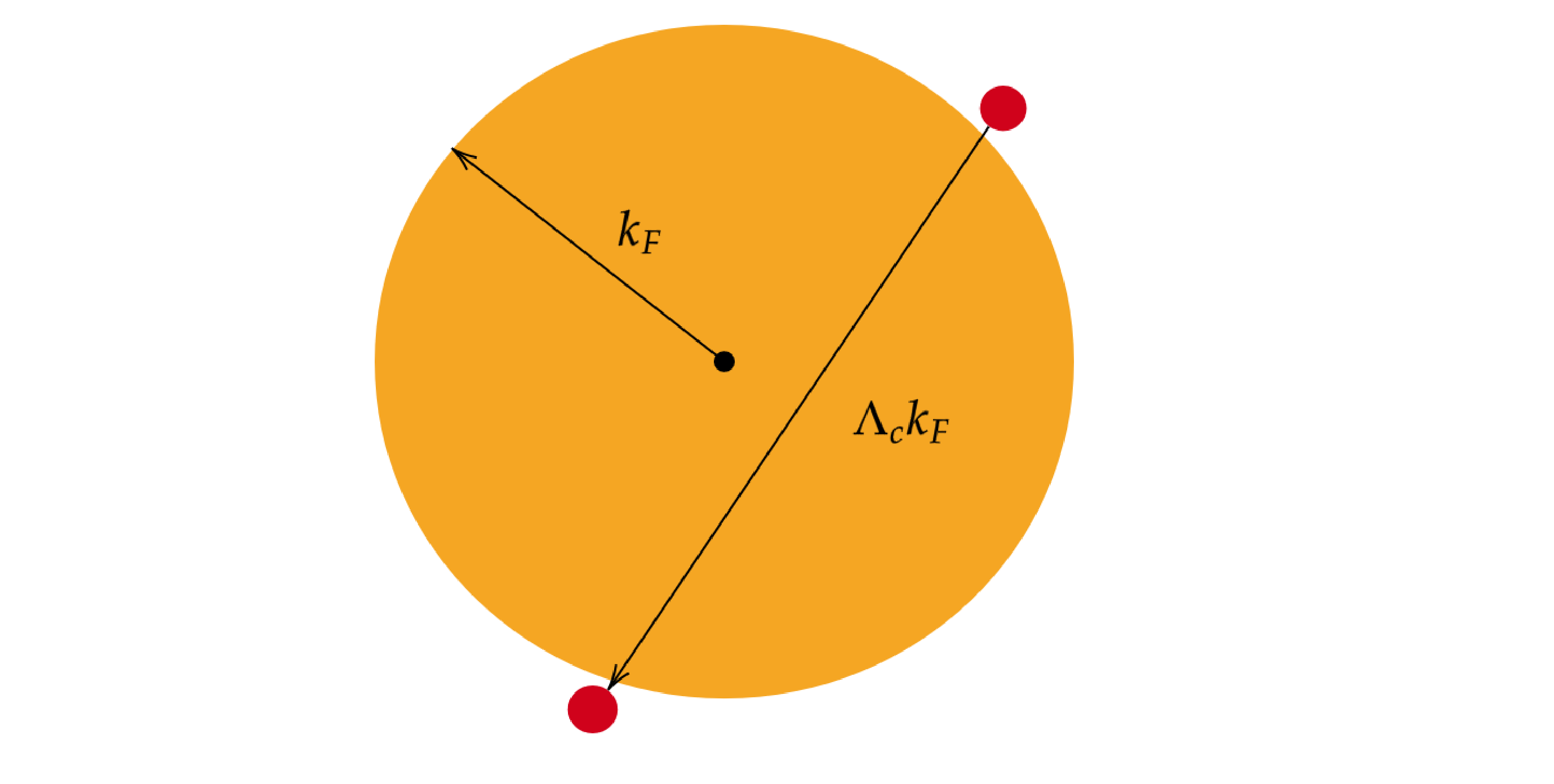}
\caption{A nucleon staying on the Fermi surface is scattered by
mesons into another point.} \label{fig:fermi}
\end{figure}

The integration range for variable $|\mathbf{q}|$ is
$|\mathbf{q}|\in [0,\Lambda_{\mathrm{c}} k^{~}_{\mathrm{F}}]$, where
$k^{~}_{\mathrm{F}}$ is the Fermi momentum and
$\Lambda_{\mathrm{c}}$ is a positive constant. For a nucleon located
precisely on the Fermi surface, it is scattered by mesons into a
different point on the surface. In principle, the maximal value of
the transferred momentum is $|\mathbf{q}|_{\mathrm{max}} =
2k^{~}_{\mathrm{F}}$. Please see Fig.~\ref{fig:fermi} for a
schematic illustration. However, the scattering process with large
momenta could be considerably suppressed by such heavy intermediate
mesons as $\omega$, $\rho$, and $\delta$. Thus, the actual upper
limit of $|\mathbf{q}|$ may take some different value from
$2k^{~}_{\mathrm{F}}$. It is therefore more convenient to take the
upper cutoff to be $|\mathbf{q}|_{\mathrm{max}} =
\Lambda_{\mathrm{c}} k^{~}_{\mathrm{F}}$, where
$\Lambda_{\mathrm{c}}$ serves as an independent tuning parameter.
The integration range for energy $\omega$ is
$[-\Omega_{\mathrm{c}},\Omega_{\mathrm{c}}]$. To include more
scattering processes, $\Omega_{\mathrm{c}}$ should be sufficiently
large. Suppose that the nucleon density is as high as
$6\rho^{~}_{\mathrm{B}0}$ in the center of neutron star core, then
the Fermi energy would be roughly $E^{~}_{\mathrm{F}} =
\sqrt{k^{2}_{\mathrm{F}} + m^{2}_{\mathrm{N}}} \approx 1000
~\mathrm{MeV}$. Thus, here we choose to take
$\Omega_{\mathrm{c}}=1000~$MeV, which appears to be large enough.
One could adopt a different value, say $\Omega_{\mathrm{c}} =
1500~$MeV. Then, all the tuning parameters would take slightly
different values, which eventually leads to only a minor change to
the final EOS of neutron stars.

To facilitate numerical computations, we re-define the internal
energy as $\omega\rightarrow \omega-\varepsilon$ and the momentum as
$\mathbf{q}\rightarrow \mathbf{q}-\mathbf{k}$, and also introduce
Matsubara energy $\varepsilon\rightarrow i\varepsilon$ and
$\omega\rightarrow i\omega$. In spherical coordinates, with
$\mathbf{q}$ direction being the axis, we have $k_{x}=|\mathbf{k}|$,
and $k_{y}=k_{z}=0$. Accordingly,
\begin{small}
\begin{eqnarray}
q_{x}=|\mathbf{q}|\sin\theta \cos\varphi, \quad q_{y} =
|\mathbf{q}|\sin\theta \sin\varphi, \quad q_{z}=
|\mathbf{q}|\cos\theta.
\end{eqnarray}
\end{small}
Then the above integral equations become
\begin{widetext}
\begin{eqnarray}
A_0(\varepsilon) &=& 1+\frac{1}{\varepsilon}\int \frac{d\omega
{|\mathbf{q}|}^{2}d|\mathbf{q}|d\cos\theta d\varphi}{(2\pi)^4}
\frac{A_0(\omega)\omega}{A^{2}_0(\omega)\omega^{2} +
A^{2}_1(\omega){|\mathbf{q}|}^{2} +A^{2}_2(\omega)m^{2}_\mathrm{N}}
\nonumber \\
&&\times\Bigg(\frac{g^{2}_{\sigma}}{(\omega-\varepsilon)^{2} +
{|\mathbf{q}|}^{2}-2|\mathbf{q}||\mathbf{k}|\sin\theta \cos\varphi +
{|\mathbf{k}|}^{2}+m^{2}_{\sigma}}-\frac{g^{2}_{\omega}}{(\omega -
\varepsilon)^{2} + {|\mathbf{q}|}^{2}-2|\mathbf{q}||\mathbf{k}|\sin\theta
\cos\varphi+{|\mathbf{k}|}^{2}+m^{2}_{\omega}} \nonumber \\
&&-\frac{g^{2}_{\rho}/4}{(\omega-\varepsilon)^{2}+{|\mathbf{q}|}^{2} -
2|\mathbf{q}||\mathbf{k}|\sin\theta\cos\varphi + {|\mathbf{k}|}^{2} +
m^{2}_{\rho}}+\frac{g^{2}_{\delta}}{(\omega-\varepsilon)^{2} +
{|\mathbf{q}|}^{2}-2|\mathbf{q}||\mathbf{k}|\sin\theta \cos\varphi +
{|\mathbf{k}|}^{2}+m^{2}_{\delta}}\Bigg), \label{eq:integraleqA00}
\nonumber \\
A_1(\varepsilon) &=& 1+\frac{1}{|\mathbf{k}|}\int \frac{d\omega
{|\mathbf{q}|}^{2}d|\mathbf{q}|d\cos\theta d\varphi}{(2\pi)^4}
\frac{A_1(\omega)|\mathbf{q}|\sin\theta \cos\varphi}{A^{2}_0(\omega)
\omega^{2}
+A^{2}_1(\omega){|\mathbf{q}|}^{2}+A^{2}_2(\omega)m^{2}_\mathrm{N}}
\nonumber \\
&&\times\Bigg(\frac{g^{2}_{\sigma}}{(\omega-\varepsilon)^{2} +
{|\mathbf{q}|}^{2}-2|\mathbf{q}||\mathbf{k}|\sin\theta \cos\varphi +
{|\mathbf{k}|}^{2}+m^{2}_{\sigma}} +
\frac{g^{2}_{\omega}}{(\omega-\varepsilon)^{2} +
{|\mathbf{q}|}^{2}-2|\mathbf{q}||\mathbf{k}|\sin\theta \cos\varphi +
{|\mathbf{k}|}^{2}+m^{2}_{\omega}} \nonumber \\
&&+\frac{g^{2}_{\rho}/4}{(\omega-\varepsilon)^{2} + {|\mathbf{q}|}^{2} -
2|\mathbf{q}||\mathbf{k}|\sin\theta \cos\varphi + {|\mathbf{k}|}^{2} +
m^{2}_{\rho}}+\frac{g^{2}_{\delta}}{(\omega-\varepsilon)^{2} +
{|\mathbf{q}|}^{2}-2|\mathbf{q}||\mathbf{k}|\sin\theta \cos\varphi +
{|\mathbf{k}|}^{2}+m^{2}_{\delta}}\Bigg),\label{eq:integraleqA11}
\nonumber \\
A_2(\varepsilon) &=& 1-\frac{1}{m_{\mathrm{N}}}\int \frac{d\omega
{|\mathbf{q}|}^{2}d|\mathbf{q}| d\cos\theta d\varphi}{(2\pi)^4}
\frac{A_2(\omega)m_{\mathrm{N}}}{A^{2}_0(\omega)\omega^{2} +
A^{2}_1(\omega)|{\mathbf{q}|}^{2} +A^{2}_2(\omega)m^{2}_\mathrm{N}} \nonumber \\
&&\times\Bigg(\frac{g^{2}_{\sigma}}{(\omega-\varepsilon)^{2} +
{|\mathbf{q}|}^{2}-2|\mathbf{q}||\mathbf{k}|\sin\theta \cos\varphi +
{|\mathbf{k}|}^{2}+m^{2}_{\sigma}}-\frac{g^{2}_{\omega}}{(\omega-\varepsilon)^{2}
+ {|\mathbf{q}|}^{2}-2|\mathbf{q}||\mathbf{k}|\sin\theta \cos\varphi +
{|\mathbf{k}|}^{2}+m^{2}_{\omega}} \nonumber \\
&&-\frac{g^{2}_{\rho}/4}{(\omega-\varepsilon)^{2} + {|\mathbf{q}|}^{2} -
2|\mathbf{q}||\mathbf{k}|\sin\theta \cos\varphi + {|\mathbf{k}|}^{2} +
m^{2}_{\rho}}+\frac{g^{2}_{\delta}}{(\omega-\varepsilon)^{2} +
{|\mathbf{q}|}^{2}-2|\mathbf{q}||\mathbf{k}|\sin\theta \cos\varphi +
{|\mathbf{k}|}^{2}+m^{2}_{\delta}}\Bigg).\label{eq:integraleqA22}
\nonumber
\end{eqnarray}
\end{widetext}
For calculational convenience, it is useful to divide all the energy
and momenta by $\Lambda_{\mathrm{c}} k^{~}_{\mathrm{F}}$.
$|\mathbf{q}|$ becomes a dimensionless variable after it is
re-scaled as
\begin{eqnarray}
|\mathbf{q}| \rightarrow \frac{|\mathbf{q}|}{ \Lambda_{\mathrm{c}}
k^{~}_{\mathrm{F}}}.
\end{eqnarray}
The re-scaled dimensionless variable $|\mathbf{q}|$ is defined in
the range $|\mathbf{q}|\in[0, 1]$. The re-scaled energy is defined
in the range $\omega
\in\left[-\frac{\Omega_{\mathrm{c}}}{\Lambda_{\mathrm{c}}
k^{~}_{\mathrm{F}}}, \frac{\Omega_{\mathrm{c}
}}{\Lambda_{\mathrm{c}} k^{~}_{\mathrm{F}}}\right]$.

These integral equations are self-consistent and should be solved by
using suitable numerical techniques. We will solve them by means of
iteration method \cite{Liu21}. First, let $A_{0,1,2}$ take some
initial values, which might be $A_0 = A_1 = A_2 = 1$. Second,
substitute the initial values into the equations of $A_{0,1,2}$ to
obtain a set of new values of $A_{0,1,2}$, which are normally
different from initial values. Third, insert these new values again
into the equations of $A_{0,1,2}$ to obtain another set of new
values of $A_{0,1,2}$. After $i$ times of iteration, one obtains
$A^i_{0}$, $A^i_{1}$, and $A^i_{2}$, which are then substituted into
the coupled equations to get $A^{i+1}_{0}$, $A^{i+1}_{1}$, and
$A^{i+1}_{2}$. Repeat such an operation many times until the
difference between $i$ results and $(i+1)$ results vanishes. The
error factors created after $i$ iterations are
\begin{eqnarray}
\mathrm{EPS}_\mathrm{0,1,2}(i) = \frac{1}{N}\sum^N_{n=1}\frac{\mid
A^i_{0,1,2}(n)-A^{i-1}_{0,1,2}(n)\mid}{\mid A^i_{0,1,2}(n)\mid+ \mid
A^{i-1}_{0,1,2}(n)\mid}.
\end{eqnarray}
For a given nucleon density $\rho^{\ast}_\mathrm{B}$,
$\mathrm{EPS}_\mathrm{0,1,2}(i)$ are found to decrease gradually
with increasing $i$. Once $\mathrm{EPS}_\mathrm{0,1,2}(i)$ become
sufficiently small, the iteration process can be terminated. Then
the finial results of $A_0$, $A_1$, and $A_2$ are determined. In
realistic calculations, convergence is believed to be achieved once
$\mathrm{EPS}_{\mathrm{0,1,2}} < 10^{-6}$. Such convergent solutions
of $A_{0,1,2}(\varepsilon)$ lead to the renormalized neutron
propagator $G(k)$.

\section{Equation of states of neutron star matter}
\label{sec:fitting}

Our next step is to incorporate quantum many-body effects into the
EOS of neutron star. In principle, one could first calculate the
Luttinger-Ward (LW) functional \cite{Luttinger} based on the
renormalized neutron propagator $G(k)$ obtained from solving the DS
equation and then use this functional to determine the EOS. Such
calculations are technically quite involved. Below, we will compute
the EOS by taking a route that is less rigorous but more
practicable.

Within the framework of RMFT \cite{Glendenningbook}, the analytical
expressions of the energy density $\epsilon$ and the pressure $P$
can be derived. This is a noticeable advantage of RMFT. We wish to
obtain the analytical expressions of $\epsilon$ and $P$ after
including the many-body effects. We find that this goal can be
achieved by performing RMFT calculations based on the following
renormalized Lagrangian density
\begin{eqnarray}
\mathcal{L}_{\mathrm{R}} &=& \bar{\psi}(i\bar{A}_{0}
\partial_{t}\gamma^{0} + i\bar{A}_1\bm{\partial}\cdot
\bm{\gamma} - \bar{A}_2 m^{~}_{\mathrm{N}})\psi \nonumber \\
&& +\frac{1}{2}\partial_{\mu}\sigma \partial^{\mu} \sigma -
\frac{1}{2}m^{2}_{\sigma}\sigma^{2} \nonumber \\
&& -\frac{1}{4}\omega_{\mu\nu}\omega^{\mu\nu} +
\frac{1}{2}m^{2}_{\omega}\omega_{\mu}\omega^{\mu} \nonumber \\
&& -\frac{1}{4}\bm{\rho}_{\mu\nu}\cdot\bm{\rho}^{\mu\nu} +
\frac{1}{2}m^{2}_{\rho} \bm{\rho}_{\mu}\cdot\bm{\rho}^{\mu} \nonumber \\
&& +\frac{1}{2}\partial_{\mu}\bm{\delta}\cdot \partial^{\mu}
\bm{\delta} - \frac{1}{2}m^{2}_{\delta} \bm{\delta}^{2}\nonumber \\
&& +g_{\sigma}\sigma\bar{\psi} \psi -
g_{\omega}\omega_{\mu}\bar{\psi}\gamma^{\mu}\psi \nonumber \\
&& -\frac{1}{2}g_{\rho}\bm{\rho}_{\mu}\cdot \bar{\psi}\bm{\tau}
\gamma^{\mu}\psi+ g_{\delta}\bm{\delta}\cdot \bar{\psi}
\bm{\tau}\psi.\label{eq:renormalizedL}
\end{eqnarray}
This Lagrangian density is written down based on the assumption that
the many-body effects are included in three constants
$\bar{A}_{0,1,2}$ that are averaged as follows
\begin{eqnarray}
{\bar A}_{0,1,2} = \frac{\int {\bar A}_{0,1,2}(\varepsilon)
d\varepsilon}{\int d\varepsilon}.
\end{eqnarray}
Although these three constants are density dependent, we assume that
$\bar{A}_{0,1,2}$ are density independent near the saturation
density. This approximation makes it much easier to perform the
partial derivative $\frac{\partial}{\partial
\rho^{\ast}_{\mathrm{B}}}$, and allows us to eventually obtain the
analytical expressions of the EOS and observable nuclear quantities.
Later, we will discuss the validity of this approximation.

We then apply the standard RMFT \cite{Glendenningbook} to deal with
$\mathcal{L}_{\mathrm{R}}$. From Eq.~(\ref{eq:renormalizedL}), it is
easy to get the equation of motion of the nucleon field
\begin{eqnarray}
&& \left[i\bar{A}_0\partial_t\gamma^0+i\bar{A}_1\bm{\partial} \cdot
\bm{\gamma}-\bar{A}_2 m_{\mathrm{N}}^{}\right]\psi(z) \nonumber \\
&=& -g_{\sigma} \sigma(z)\psi(z)+ g_{\omega}\omega_{\mu}(z)
\gamma^{\mu}\psi(z) \nonumber \\
&& +\frac{g_{\rho}}{2}\bm{\rho}_{\mu}(z) \cdot
\bm{\tau}\gamma^{\mu}\psi(z)-g_{\delta}\bm{\delta}(z) \cdot
\bm{\tau}\psi(z).\label{eq:eomnucleon}
\end{eqnarray}
The equations of motion of the meson fields are
\begin{eqnarray}
\left(\partial_{\mu}\partial^{\mu}+m^{2}_{\sigma}\right)\sigma(z)
&=& g_{\sigma}\bar{\psi}(z)\psi(z), \label{eq:sigmaeom}\\
\partial_{\mu}\omega^{\mu\nu}(z)+m^{2}_{\omega}\omega^{\nu}(z) &=&
g_{\omega}\bar{\psi}(z)\gamma^{\nu}\psi(z), \label{eq:omegaeom}\\
\partial_{\mu}\bm{\rho}^{\mu\nu}(z)+m^{2}_{\rho}\bm{\rho}^{\nu}(z) &=&
\frac{g_{\rho}}{2}\bar{\psi}(z)\bm{\tau}\gamma^{\nu}\psi(z)
\nonumber \\
&& +g_{\rho} \bm{\rho}_{\mu}(z)\times\bm{\rho}^{\mu\nu}(z),
\label{eq:rhoeom} \\
\left(\partial_{\mu}\partial^{\mu}+m^{2}_{\delta}\right)
\bm{\delta}(z) &=& g_{\delta}\bar{\psi}(z) \bm{\tau}\psi(z).
\label{eq:deltaeom}
\end{eqnarray}
At mean-field level, all the meson fields are substituted by their
mean values, namely
\begin{eqnarray}
\sigma(z) &\rightarrow& \langle\sigma(z)\rangle=\sigma, \label{sigammean}\\
\omega_{\mu}(z) &\rightarrow& \langle\omega_{\mu}(z)\rangle =
\omega_0, \label{omegamean}\\
\bm{\rho}_{\mu}(z) &\rightarrow& \langle\bm{\rho}_{\mu}(z)\rangle =
\rho_{0(3)},\label{rhomean}\\
\bm{\delta}(z) &\rightarrow& \langle\bm{\delta}(z)\rangle =
\delta_{3}.\label{deltamean}
\end{eqnarray}
Then the renormalized Lagrangian density given by
Eq.~(\ref{eq:renormalizedL}) becomes
\begin{eqnarray}
\mathcal{L}_{\mathrm{R}}^{\mathrm{MF}} &=& \bar{\psi}(z)
(i\bar{A}_0\partial_t\gamma^0 + i\bar{A}_1\bm{\partial}\cdot
\bm{\gamma} - \bar{A}_2 m_{\mathrm{N}}^{})\psi(z) \nonumber \\
&& -\frac{1}{2}m^{2}_{\sigma}\sigma^{2} +
\frac{1}{2}m_{\omega}^{2}\omega^{2}_0 +
\frac{1}{2}m_{\rho}^{2}\rho^{2}_{0(3)} -
\frac{1}{2}m^{2}_{\delta}\delta_3^{2}\nonumber \\
&& +g_{\sigma}\sigma\bar{\psi}(z)\psi(z)-g_{\omega}\omega_0
\bar{\psi}(z) \gamma^0\psi(z) \nonumber \\
&& -\frac{g_{\rho}}{2} \rho_{0(3)}\bar{\psi}(z)\tau_3
\gamma^0\psi(z)\nonumber \\
&& +g_{\delta}\delta_{3}\bar{\psi}(z)\tau_3\psi(z).
\label{eq:RMFL}
\end{eqnarray}
Then Eq.~(\ref{eq:eomnucleon}) is simplified to
\begin{eqnarray}
&&\big[i\bar{A}_0\partial_t\gamma^0+i\bar{A}_1\bm{\partial}\cdot
\bm{\gamma}-g_{\omega}\omega_0\gamma^0-\frac{g_{\rho}}{2}
\rho_{0(3)}\tau_3\gamma^{0} \nonumber
\\
&& -(\bar{A}_2 m_{\mathrm{N}}^{}-g_{\sigma}\sigma -
g_{\delta}\delta_3\tau_3)\big]\psi(z)=0, \label{fermionRMFeom}
\end{eqnarray}
and Eqs.~(\ref{eq:sigmaeom})-(\ref{eq:deltaeom}) are converted to
\begin{eqnarray}
\sigma &=& \frac{g_{\sigma}}{m^{2}_{\sigma}}\langle\bar{\psi}(z)
\psi(z)\rangle \nonumber \\
&=& \frac{g_{\sigma}}{m^{\ast2}_{\sigma}} \rho^{*}_\mathrm{s},
\label{sigmaRMFeom} \\
\omega^{0} &=& \frac{g_{\omega}}{m^{2}_{\omega}}\langle\bar{\psi}(z)
\gamma^{0}\psi(z)\rangle = \frac{g_{\omega}}{m^{2}_{\omega}}\langle
\psi^{\dag}(z)\psi(z)\rangle \nonumber \\
&=& \frac{g_{\omega}}{m^{2}_{\omega}}\rho^{*}_\mathrm{B},
\label{omegaRMFeom} \\
\rho_{3}^{0} &=& \frac{g_{\rho}}{2m^{2}_{\rho}} \langle
\bar{\psi}(z)\tau_3 \gamma^0\psi(z)\rangle = \frac{g_{\rho}}{2
m^{2}_{\rho}}\langle\psi^\dag(z) \tau_3\psi(z)\rangle \nonumber \\
&=& \frac{g_{\rho}}{2m^{2}_{\rho}}\rho^{*}_3,
\label{rhoRMFeom} \\
\delta_{3} &=& \frac{g_\delta}{m^{2}_{\delta}} \langle \bar{\psi}(z)
\tau_3\psi(z)\rangle \nonumber \\
&=& \frac{g_{\delta}}{m^{2}_{\delta}}\rho^{*}_{\mathrm{s}3}.
\label{deltaRMFeom}
\end{eqnarray}
Here, a raised asterisk is used to denote the inclusion of quantum
many-body effects. The baryon density is $\rho^{\ast}_\mathrm{B} =
\rho^{\ast}_\mathrm{p} + \rho^{\ast}_\mathrm{n}$, where
\begin{eqnarray}
\rho^{\ast}_i = 2\int^{k_{\mathrm{F},i}}_{0}
\frac{d^3\mathbf{k}}{(2\pi)^{3}}\frac{1}{\bar{A}_{0,i}}
\end{eqnarray}
is the renormalized baryon density. The renormalized scalar density
is $\rho^{\ast}_\mathrm{s} = \rho^{\ast}_{\mathrm{s,p}} +
\rho^{\ast}_{\mathrm{s,n}}$, where
\begin{eqnarray}
\rho^{\ast}_{\mathrm{s},i} &=& 2\int^{k_{\mathrm{F},i}}_0
\frac{d^3\mathbf{k}}{(2\pi)^3}
\frac{m^{\ast}_{\mathrm{N}}/\bar{A}_{0,i}}{E^{\ast}_{\mathrm{F},
i}(\mathbf{k})},\\
E^{\ast}_{\mathrm{F},i}(\mathbf{k}) &=&
\sqrt{\frac{\bar{A}^{2}_{1,i}}{\bar{A}^{2}_{0,i}}\mathbf{k}^{2}+
m^{\ast2}_{\mathrm{N},i}},\\
m^\ast_{\mathrm{N},i} &=& \frac{\bar{A}_{2,i} m_{\mathrm{N},i} -
\frac{g^{2}_{\sigma}}{m^{\ast2}_{\sigma}}\rho^{*}_\mathrm{s}-
\frac{g^{2}_{\delta}}{m^{2}_{\delta}}\rho^{*}_{\mathrm{s}3}
\tau_3}{\bar{A}_{0,i}},
\end{eqnarray}
and we have replaced $m_\sigma$ with the renormalized $\sigma$ meson
mass $m^\ast_{\sigma}$. The renormalized isospin density is
\begin{eqnarray}
\rho^{\ast}_3 = \rho^{\ast}_{\mathrm{p}} - \rho^{\ast}_{\mathrm{n}}
= (2y-1)\rho^{\ast}_\mathrm{B},
\end{eqnarray}
where the proton fraction is $y =
\rho^{\ast}_{\mathrm{p}}/\rho^{\ast}_\mathrm{B}$. The renormalized
isospin scalar density is
\begin{eqnarray}
\rho^{\ast}_{\mathrm{s}3} = \rho^{\ast}_{\mathrm{s,p}} -
\rho^{\ast}_{\mathrm{s,n}}.
\end{eqnarray}

It is emphasized that the above results are applicable to both
symmetric and asymmetric nuclear matter. The fraction of neutrons
can be easily tuned by changing the value of $y$.

\subsection{EOS and observable quantities}
\label{sec:subseceos}

The expectation value of the energy-momentum tensor in the rest
frame of the matter is diagonal, namely
\begin{eqnarray}
\langle T^{\mu\nu}\rangle = \mathrm{diag}(\epsilon,P,P,P).
\end{eqnarray}
Then the energy density $\epsilon=\langle T^{00}\rangle$ and the
pressure $P=\frac{1}{3}\langle T^{ii}\rangle$ of the system are
given by
\begin{eqnarray}
\epsilon &=& \sum_{i=p,n}2\int^{k_{\mathrm{F},i}}_0
\frac{d^3\mathbf{k}}{(2\pi)^3}E^\ast_{\mathrm{F},i}(\mathbf{k})
+\frac{g^{2}_{\sigma}}{2m^{\ast 2}_{\sigma}}\rho^{\ast
2}_{\mathrm{s}} + \frac{g^{2}_{\omega}}{2m^{2}_{\omega}}
\rho^{\ast2}_{\mathrm{B}} \nonumber \\
&& + \frac{g^{2}_{\rho}}{8m^{2}_{\rho}}\rho^{\ast2}_3 +
\frac{g^2_{\delta}}{2m^2_{\delta}}\rho^{\ast2}_{\mathrm{s}3},
\label{energyappendix} \\
P &=& \sum_{i=p,n}\frac{2}{3}\int^{k_{\mathrm{F},i}}_0
\frac{d^3\mathbf{k}}{(2\pi)^3}\frac{\bar{A}^{2}_{1,i}
\mathbf{k}^{2}/\bar{A}^{2}_{0,i}}{E^\ast_{\mathrm{F},i}(\mathbf{k})}
- \frac{g^{2}_{\sigma}}{2m^{\ast 2}_{\sigma}}
\rho^{\ast2}_{\mathrm{s}} \nonumber \\
&& +\frac{g^{2}_{\omega}}{2m^{2}_{\omega}}\rho^{\ast2}_{\mathrm{B}}
+ \frac{g^{2}_{\rho}}{8m^{2}_{\rho}}\rho^{\ast2}_3 -
\frac{g^2_{\delta}}{2m^2_{\delta}}\rho^{\ast2}_{\mathrm{s}3}.
\label{presureappendix}
\end{eqnarray}

We have verified that the thermodynamic relationship
$P=\rho^{\ast2}_\mathrm{B} \frac{\partial
(\epsilon/\rho^{\ast}_\mathrm{B})}{\partial\rho^{\ast}_\mathrm{B}}$
is satisfied. The above expressions of $\epsilon$ and $P$ are then
used to compute the binding energy $E^{}_\mathrm{b}$, the effective
nucleon mass $m^{{\ast}}_{\mathrm{N}}$, the compressible modulus
$K$, and the symmetry energy $E_{\mathrm{s}}$ at the saturation
density of symmetric nuclear matter as follows
\begin{eqnarray}
E_\mathrm{b} &=& \frac{\epsilon}{\rho^{{\ast}}_\mathrm{B}} -
m_{\mathrm{N}}^{}, \label{bindingenergy}\\
m^{{\ast}}_{\mathrm{N}} &=& \frac{\bar{A}_2 m_{\mathrm{N}}^{} -
\frac{g^{2}_{\sigma}}{m^{\ast2}_{\sigma}}
\rho^{\ast}_\mathrm{s}}{\bar{A}_0},
\label{effectivemass}\\
K &=& 9 \rho^{\ast2}_\mathrm{B} \frac{\partial^{2}
(\epsilon/\rho^{\ast}_\mathrm{B})}{\partial \rho^{\ast2}_\mathrm{B}}
\nonumber \\
&=& \frac{3 \frac{\bar{A}^{2}_1}{\bar{A}_0} k^{2}_{\mathrm{F}}
}{E^{\ast}_\mathrm{F}}+\frac{9g^{2}_{\omega}
\rho^{\ast}_\mathrm{B}}{m^{2}_{\omega}}-\frac{9g^{2}_{\sigma}
\frac{m^{{\ast} 2}_\mathrm{N}}{E^{{\ast}2}_\mathrm{F}}
\rho^{\ast}_\mathrm{B}}{m^{{\ast}2}_{\sigma} +
\frac{3g^{2}_{\sigma}}{\bar{A}_0}
(\frac{\rho^{*}_\mathrm{s}}{m_{\mathrm{N}}^{\ast}} -
\frac{\rho^{*}_\mathrm{B}}{E^{\ast}_\mathrm{F}})},
\label{compressionmodulus}\\
E_{\mathrm{s}} &=& \frac{1}{8}\frac{\partial^{2}
(\epsilon/\rho^\ast_\mathrm{B})}{\partial y^{2}} \nonumber \\
&=& \frac{\frac{\bar{A}^{2}_1}{\bar{A}_0} k^{2}_\mathrm{F}}{6
E^{\ast}_\mathrm{F}}+\frac{g^{2}_{\rho} \rho^{{\ast}}_\mathrm{B}
}{8m^{2}_{\rho}}-\frac{g^{2}_{\delta}\frac{m^{\ast
2}_\mathrm{\mathrm{N}}}{2E^{\ast2}_\mathrm{F}}\rho^\ast_\mathrm{B}}
{m^2_\delta+\frac{3g^{2}_{\delta}}{\bar{A}_0}
(\frac{\rho^{*}_\mathrm{s}}{m_{\mathrm{N}}^{\ast}} -
\frac{\rho^{*}_\mathrm{B}}{E^{\ast}_\mathrm{F}})}.
\label{symmetryenergy}
\end{eqnarray}
The symmetry energy slope $L_{\mathrm{s}}$ is found to have the
following complicated analytical expression:
\begin{widetext}
\begin{eqnarray}
L_{\mathrm{s}} &=& 3\rho^{*}_\mathrm{B}\frac{\partial
E_{\mathrm{s}}}{\partial \rho^{*}_\mathrm{B}} \nonumber \\
&=&\frac{\frac{\bar{A}^{2}_1}{\bar{A}_0} k^{2}_\mathrm{F}}{3
E^{\ast}_\mathrm{F}} + \frac{3g^{2}_{\rho}
\rho^{{\ast}}_\mathrm{B}}{8m^{2}_{\rho}} -
\frac{\frac{\bar{A}^{4}_1}{\bar{A}^{3}_0} k^{4}_\mathrm{F}}{6
E^{\ast3}_\mathrm{F}}\left(1-\frac{\frac{
g^{2}_{\sigma}}{\bar{A}^{2}_1} \frac{2m^{{\ast} 2}_\mathrm{N}}{
\pi^{2} E^{{\ast}}_\mathrm{F}} k_\mathrm{F}}{m^{{\ast}2}_{\sigma} +
\frac{3 g^{2}_{\sigma}}{\bar{A}_0}
(\frac{\rho^{*}_\mathrm{s}}{m_{\mathrm{N}}^{\ast}} -
\frac{\rho^{*}_\mathrm{B}}{E^{\ast}_\mathrm{F}})}\right)\nonumber\\
&& - \frac{g^{2}_{\delta}\frac{m^{\ast 2}_\mathrm{\mathrm{N}}}{2
E^{\ast 2}_\mathrm{F}} \rho^{\ast}_{\mathrm{B}}}{m^{2}_{\delta} +
\frac{3g^{2}_{\delta}}{\bar{A}_0}
(\frac{\rho^{*}_\mathrm{s}}{m_{\mathrm{N}}^{\ast}} -
\frac{\rho^{*}_\mathrm{B}}{E^{\ast}_\mathrm{F}})}
\left\{3-\frac{2\frac{\bar{A}^{2}_1}{\bar{A}^{2}_0}
k^{2}_\mathrm{F}}{E^{\ast2}_\mathrm{F}}-\frac{6
\frac{g^{2}_{\sigma}}{\bar{A}_0} \Big(
\frac{1}{E^{{\ast}}_\mathrm{F}}-\frac{m_{\mathrm{N}}^{\ast
2}}{E^{\ast3}_\mathrm{F}}\Big)\rho^{\ast}_\mathrm{B}}{m^{{\ast}
2}_{\sigma} + \frac{3g^{2}_{\sigma}}{\bar{A}_0}
(\frac{\rho^{*}_\mathrm{s}}{m_{\mathrm{N}}^{\ast}} -
\frac{\rho^{*}_\mathrm{B}}{E^{\ast}_\mathrm{F}})}
+\frac{3g^{2}_{\delta} \rho^{\ast}_{\mathrm{B}}}{m^{{\ast}
2}_{\delta} + \frac{3 g^{2}_{\delta}}{\bar{A}_0}
(\frac{\rho^{*}_\mathrm{s}}{m_{\mathrm{N}}^{\ast}} -
\frac{\rho^{*}_\mathrm{B}}{E^{\ast}_\mathrm{F}})}\right.\nonumber\\
&& \left.\times\left[\frac{6 \frac{g^{2}_{\sigma}}{\bar{A}^{2}_0}
\frac{1}{E^{{\ast}}_\mathrm{F}}(\frac{\rho^{*}_\mathrm{s}}
{m_{\mathrm{N}}^{\ast}} -
\frac{\rho^{*}_\mathrm{B}}{E^{\ast}_\mathrm{F}})
}{m^{{\ast}2}_{\sigma} + \frac{3g^{2}_{\sigma}}{\bar{A}_0}
(\frac{\rho^{*}_\mathrm{s}}{m_{\mathrm{N}}^{\ast}} -
\frac{\rho^{*}_\mathrm{B}}{E^{\ast}_\mathrm{F}})} -
\frac{\frac{\bar{A}^{2}_1}{\bar{A}^{3}_0}
k^{2}_\mathrm{F}}{E^{\ast3}_\mathrm{F}}\left(1+\frac{3
\frac{g^{2}_{\sigma}}{\bar{A}_0}
\frac{\rho^{*}_\mathrm{B}}{E^{{\ast}}_\mathrm{F}}
}{m^{{\ast}2}_{\sigma} + \frac{3g^{2}_{\sigma}}{\bar{A}_0}
(\frac{\rho^{*}_\mathrm{s}}{m_{\mathrm{N}}^{\ast}} -
\frac{\rho^{*}_\mathrm{B}}{E^{\ast}_\mathrm{F}})}\right)\right]\right\}.
\label{symmetryenergyslope}
\end{eqnarray}
\end{widetext}

\begin{figure}[htbp]
%\centering
\includegraphics[width=3.68in]{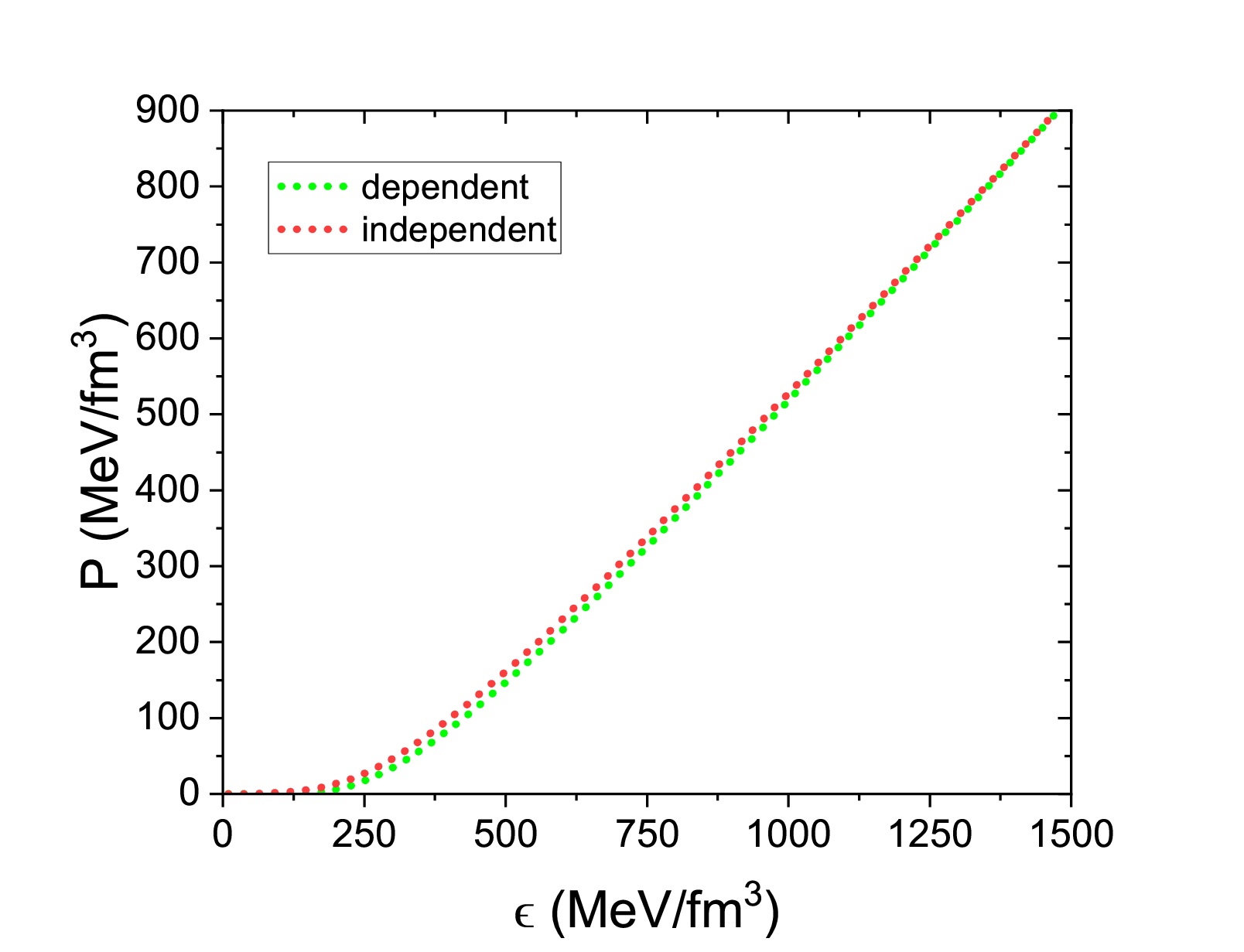}
\caption{Comparing the EOS of a pure neutron matter generated by
including the density dependence of $\bar{A}_{0,1,2}$ to the one
obtained by ignoring the density dependence of $\bar{A}_{0,1,2}$.}
\label{fig:densitydependence}
\end{figure}

The theoretical results $\rho^{}_{\mathrm{B}0}$, $E_{\mathrm{b}}$,
$m^{\ast}_{\mathrm{N}}$, $K$, $E_{\mathrm{s}}$, and $L_{\mathrm{s}}$
depend on tuning parameters $g_{\sigma}$, $g_{\omega}$, $g_{\rho}$,
$g_{\delta}$, $m^{\ast}_{\sigma}/m_{\sigma}$, and
$\Lambda_{\mathrm{c}}$. We require such tuning parameters to take
all the possible values, exploiting the Markov chain Monte Carlo
\cite{MCMC} method, until the theoretical results of these
quantities are compatible with the experimental data:
$\rho^{~}_{\mathrm{B0}}=(0.16\pm0.01)~\mathrm{fm}^{-3}$,
$E_{\mathrm{b}}=(-16\pm1)~\mathrm{MeV}$,
$m^{\ast}_{\mathrm{N}}/m^{~}_{\mathrm{N}}=(0.56-0.75)$, $K=(240\pm
20)~\mathrm{MeV}$, $E_{\mathrm{s}}=(28-34)~\mathrm{MeV}$, and
$L_{\mathrm{s}}=(40-68)~\mathrm{MeV}$.

By adjusting $g_{\sigma}$, $g_{\omega}$ and $g_{\rho}$, previous
RMFT studies of the $\sigma$-$\omega$-$\rho$ model
\cite{Glendenningbook} have yielded $\rho^{~}_{\mathrm{B0}}=0.153
~\mathrm{fm}^{-3}$, $E_\mathrm{b}=-16.3~\mathrm{MeV}$,
$m^{\ast}_{\mathrm{N}}/m^{~}_{\mathrm{N}}=0.50$,
$K=550~\mathrm{MeV}$, and $E_{\mathrm{s}}=32.5~\mathrm{MeV}$.
Apparently, $m^{\ast}_{\mathrm{N}}$ is too small and $K$ is too
large. To obtain larger $m^{\ast}_{\mathrm{N}}$ and smaller $K$, it
is necessary to introduce more tuning parameters. In RMFT, this
problem is solved by manually including some nonlinear meson
coupling terms, such as $-b_{0}\sigma^{3}$, $-c_{0}\sigma^{4}$, and
$-d_{0}\sigma^{2}\omega^{2}$. As mentioned in
Sec.~\ref{sec:introduction}, such nonlinear terms might drive the
system to become unstable. We find it sufficient to tune one single
parameter: the ratio $m^{\ast}_{\sigma}/m_{\sigma}$.

The above analytical results were obtained by ignoring the density
dependence of $\bar{A}_{0,1,2}$. It is straightforward to take into
account the density dependence of $\bar{A}_{0,1,2}$. But then it
becomes impossible to derive the analytical expressions of these
quantities. We performed direct numerical calculations based on
density dependent $\bar{A}_{0,1,2}$ and show the results in
Fig.~\ref{fig:densitydependence}. One can see that the inclusion of
the density dependence of $\bar{A}_{0,1,2}$ only slightly modifies
the final EOS. Therefore, it is well justified to drop the density
dependence of $\bar{A}_{0,1,2}$ in the calculation of the above
observable quantities.

\subsection{$\beta$ equilibrium}\label{sec:betaequilibrium}

While a pure neutron matter fulfils the charge neutrality
requirement, it cannot be stable. In the cores of neutron stars,
there are also, in addition to neutrons, a small fraction of protons
and electrons. As the density $\rho^{\ast}_{\mathrm{B}}$ increases,
some electrons are replaced by muons when the Fermi energy of
electrons surpasses the rest energy of muons, which is energetically
more favorable. The conditions of chemical equilibrium
\cite{Glendenningbook} are
\begin{eqnarray}
\mu_{p}+\mu_{e} &=& \mu_{n},\label{chemcon1}\\
\mu_{e} &=& \mu_{\mu},\label{chemcon2}
\end{eqnarray}
where the chemical potentials of neutrons, protons, electrons, and
muons are given by
\begin{eqnarray}
\mu_{n} &=& \sqrt{\frac{\bar{A}^2_{1,n}}{\bar{A}^2_{0,n}} k_{n}^{2}
+ m^{\ast2}_{\mathrm{N},n}}+\frac{g^{2}_{\omega}}{m^{2}_{\omega}}
\rho^{\ast}_{B} - \frac{g^{2}_{\rho}}{4m^{2}_{\rho}}
\rho^{\ast}_{3}, \\
\mu_{p} &=& \sqrt{\frac{\bar{A}^{2}_{1,p}}{\bar{A}^{2}_{0,p}}
k_{p}^{2} + m^{\ast 2}_{\mathrm{N},p}} +
\frac{g^{2}_{\omega}}{m^{2}_{\omega}} \rho^{\ast}_{B} +
\frac{g^{2}_{\rho}}{4m^{2}_{\rho}}\rho^{\ast}_{3}, \\
\mu_{e} &=& \sqrt{k^{2}_{e}+m^{2}_{e}}, \\
\mu_{\mu} &=& \sqrt{k^{2}_{\mu}+m^{2}_{\mu}}.
\end{eqnarray}
Here, $k_{n}$, $k_{p}$, $k_{e}$, and $k_{\mu}$ denote the Fermi
momenta of neutrons, protons, electrons, and muons, respectively.
Notice that the quantum many-body effects are already included in
these chemical potentials. The rest lepton masses are $m_{e} =
0.511~$MeV and $m_{\mu} = 105.7~$MeV. The lepton densities are
related to the corresponding Fermi momenta via the relations
$\rho_{e,\mu}=k^3_{e,\mu}/(3\pi^2)$. Also, the neutron-star core
should preserve the baryon number conservation and ensure the
electric neutrality given by two equalities:
\begin{eqnarray}
\rho^{\ast}_{\mathrm{B}} &=& \rho_{n}+\rho_{p},\label{barcon}\\
Q_{p}+Q_{e}+Q_{\mu} &=& \rho_{p}-\rho_{e}-\rho_{\mu} = 0.
\label{neucon}
\end{eqnarray}
Here, $Q_{p}$, $Q_{e}$, and $Q_{\mu}$ are the electric charges
carried by proton, electron, and muon, respectively. Making use of
Eqs.~(\ref{chemcon1}-\ref{neucon}), we determine the densities of
neutrons, protons, electrons, and muons at a given total density
$\rho^{\ast}_{\mathrm{B}}$, which then generates a more realistic
EOS for neutron stars that satisfies the $\beta$ equilibrium:
\begin{eqnarray}
\epsilon_{\beta} &=& \epsilon+\sum_{l=e,\mu}
\frac{1}{\pi^2}\int^{k_l}_0 k^2 dk
\sqrt{k^2_l+m^2_l},\label{energyBATA} \\
P_{\beta} &=& P+\sum_{l=e,\mu}\frac{1}{3\pi^2}\int^{k_{l}}_{0} dk
\frac{k^{4}}{\sqrt{k^{2}+m^{2}_{l}}}. \label{presureBATA}
\end{eqnarray}

\section{Mass-radius relation, tidal deformability, and sound speed}
\label{sec:eosofnss}

The relation between the mass $M$ and radius $R$ of neutron stars
can be obtained by inserting the EOS into the Tolman-Oppenheimer-Volkoff (TOV) equations
\cite{Tolman34, Oppenheimer39} given by
\begin{eqnarray}
\frac{dP(r)}{dr} &=& -\frac{[P(r)+\epsilon(r)][M(r)+4\pi
r^3P(r)]}{r[r-2M(r)]}, \\
\frac{dM(r)}{dr} &=& 4\pi r^{2}\epsilon(r).\label{eq:tovequation}
\end{eqnarray}
Here, $M(r)$ represents the gravitational mass enclosed within a
sphere of radius $r$ and $\epsilon(r)$ and $P(r)$ are the energy
density and the pressure at $r$, respectively.

%\begin{widetext}
\begin{table*}[htbp]\scriptsize
\caption{Simulated model parameters and nuclear quantities computed
at saturation density.}
\begin{threeparttable}
\begin{supertabular}{cccccccc}
\hline\hline $\bm{\mathrm{Model}}$ & $\bm{g_\sigma}$ &
$\bm{g_\omega}$ & $\bm{g_\rho}$ & $\bm{g_\delta}$ &
$\bm{m^\ast_\sigma/m^{~}_\sigma}$ &
$\bm{\Lambda_{\mathrm{c}}}$  \\
\hline $\sigma\omega$ & $11.2852$ & $11.4243$ & $0.0$ & $0.0$ &
$1.2204$ & $2.4672$
\\ $\sigma\omega\rho1$ & $16.8355$ &
$10.9381$ & $8.0028$ & $0.0$ & $1.9501$ & $2.0$ \\
$\sigma\omega\rho2$ & $17.5758$ & $11.3153$ & $6.5393$ & $0.0$
& $1.9355$ & $1.6011$ \\
$\sigma\omega\rho\delta$ & $17.5758$ & $11.3153$ & $6.5393$ &
$1.3554$ & $1.9355$ & $1.6011$
%$\sigma\omega\rho\delta2$ & $17.5758$ & $11.3153$ & $6.5393$ &
%$2.3454$ & $1.9355$ & $1.6011$
\\ %$\sigma\omega\rho\delta3$ & $17.575787$ &
%$11.315254$ & $6.539279$ & $3.065440$ & $1.935519$ & $1.601081$
%\\
%$\sigma\omega\rho\delta4$ & $17.575787$ & $11.315254$ & $6.539279$ &
%$4.355440$ & $1.935519$ & $1.601081$\\\hline
& $\bm{\rho^{~}_{\mathrm{B0}}}$ & $\bm{E_\mathrm{b}}$ &
$\bm{m^{\ast}_\mathrm{N}}$ & $\bm{K}$ & $\bm{E_{\mathrm{s}}}$ &
$\bm{L_{\mathrm{s}}}$\\
& $(\mathrm{fm}^{-3})$ & $(\mathrm{MeV})$ & $(m^{~}_\mathrm{N})$ &
$(\mathrm{MeV})$ & $(\mathrm{MeV})$ & $(\mathrm{MeV})$\\
\hline $\sigma\omega$ & $ 0.1580$ & $-16.4176$ & $0.6327$ &
$250.1615$ & $19.0599$ & $57.8220$\\
$\sigma\omega\rho1$  & $ 0.1599$ & $ -16.8985$ & $0.6699$ &
$260.6649$ &
$34.3386$& $101.4963$\\
$\sigma\omega\rho2$ & $0.1597$ & $-16.4255$
& $0.6462$ & $220.5497$ & $28.9727$ & $87.5612$\\
$\sigma\omega\rho\delta$ & $0.1597$ & $-16.8108$ & $0.6457$ &
$221.3642$ & $28.0002$ & $85.4057$\\
%$\sigma\omega\rho\delta2$ &
%$0.1596$ & $-17.5793$ & $0.6448$ & $222.9915$ &
%$26.0646$ & $81.1394$\\
%$\sigma\omega\rho\delta3$  & $0.159604$ &
%$-18.397031$ & $0.643880$ & $224.726586$ & $24.012453$ & $76.649981$\\
%$\sigma\omega\rho\delta4$ & $0.159500$ & $-20.408471$ & $0.641518$ &
%$229.011212$ & $18.996035$ & $65.825229$\\
\hline\hline
\end{supertabular}\label{TAB1}
%\begin{tablenotes}\footnotesize
%\item[]Experimental values of these quantities: $\rho^{~}_{\mathrm{B0}} =
%(0.16\pm0.01)\mathrm{fm}^{-3}$, $E_{\mathrm{b}} =
%(-16\pm1)\mathrm{MeV}$ $m^{\ast}_{\mathrm{N}}/m^{~}_{\mathrm{N}} =
%(0.56\sim 0.75)$, $K=(240\pm 20)\mathrm{MeV}$, and
%$E_{\mathrm{s}}=(28\sim 34)\mathrm{MeV}$.
%\end{tablenotes}
\end{threeparttable}
\end{table*}
%\end{widetext}

In addition to the $M$-$R$ relation, tidal deformability is another
important observable quantity that is frequently used to probe the
interior structure of neutron stars. This quantity measures the
deformation of a neutron star under the influence of an external
gravitational field exerted by other stars. The tidal deformability
can be quantified by a dimensionless parameter \cite{Hinderer08,
Postnikov10}:
\begin{eqnarray}
\Lambda = \frac{2}{3}k_{2}C^{-5},\label{eq:tidaldeformability}
\end{eqnarray}
where $C = M/R$ is the compactness parameter and $k_{2}$, related to
the neutron star EOS, refers to the second-order Love number.
Detailed calculations and derivations of $k_{2}$ can be found in
Refs.~\cite{Hinderer08, Postnikov10}.

\begin{table*}[htbp]\scriptsize
\centering \caption{Dependence of $L_{\mathrm{s}}$ (at saturation
density) and $\Lambda_{\mathrm{1.4}}$ on the tuning parameter
$a_\rho$ for the $\sigma\omega\rho2$ model.}
\begin{threeparttable}
\begin{supertabular}{cccccccccccc}
\hline\hline $\bm{\mathrm{Model}}$ & $ L87.56$ & $L80$ & $L70$ &
$L60$ & $L50$ &
$L40$ & $L30$\\
\hline $\bm{a_\rho}$ & $ 0$ & $0.1123$ & $0.2610$& $0.4097$ &
$0.5583$&
$0.7070$ & $0.8557$\\
$\bm{L_{\mathrm{s}}}(\mathrm{MeV})$  & $ 87.56$ & $80$ & $70$& $60$
& $50$ & $40$ &
$30$\\
$\bm{\Lambda_{\mathrm{1.4}}}$ & $1225$ & $1112$ & $1038$
& $954$ & $903$& $872$ & $847$\\
\hline\hline
\end{supertabular}
\label{TAB2}
%\begin{tablenotes}\footnotesize
%\item[]
%\end{tablenotes}
\end{threeparttable}
\end{table*}

After carrying out extensive calculations, we find a few suitable
sets of parameters and show them in Table \ref{TAB1}. Among these
models, the $\sigma\omega\rho2$ model appears to be an ideal
candidate. The values of $\rho^{}_{\mathrm{B}0}$, $E_{\mathrm{b}}$,
$m^{\ast}_{\mathrm{N}}$, $K$, and $E_{\mathrm{s}}$ produced by this
model are very consistent with experiments. But the slope
$L_{\mathrm{s}}=87.56~\mathrm{MeV}$ seems too large. Accordingly, the
tidal deformability $\Lambda_{1.4}$ for a neutron star with mass
$1.4~M_{\odot}$ is also unrealistically large.

\begin{figure}[htbp]
\centering
\includegraphics[width=3.7in]{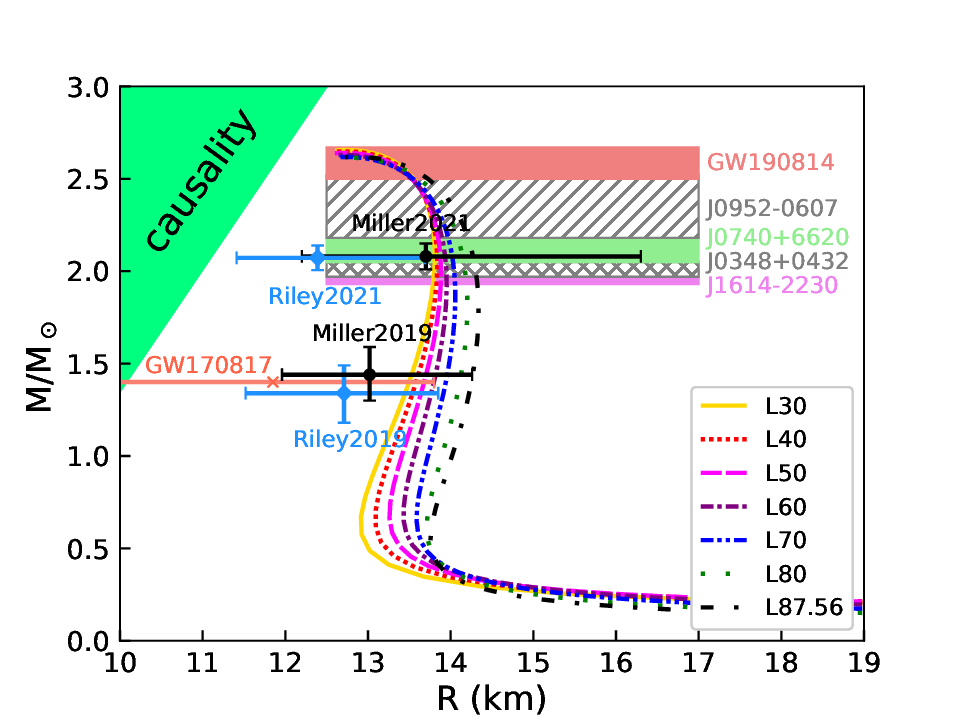}
\caption{Comparison between the theoretical results of the $M$-$R$
relations obtained from $\sigma\omega\rho2$ model and astrophysical
observations of neutron stars.} \label{fig:mr}
\end{figure}

To get more realistic $L_{\mathrm{s}}$ and $\Lambda_{1.4}$, we
replace constant $g_{\rho}$ by a density dependent function
\cite{TW99, Shen21} given by
\begin{eqnarray}
\Gamma_{\rho} = g_{\rho} \exp\left[-a_{\rho}
\left(\frac{\rho^{\ast}_{\mathrm{B}}}{ \rho^{~}_{\mathrm{B0}}} -
1\right)\right],
\end{eqnarray}
where $a_{\rho}$ is the sixth tuning parameter for the
$\sigma\omega\rho2$ model. As $a_{\rho}$ grows from zero, the values
of $\rho^{}_{\mathrm{B}0}$, $E_{\mathrm{b}}$,
$m^{\ast}_{\mathrm{N}}$, $K$, and $E_{\mathrm{s}}$ listed in Table
\ref{TAB1} remain unchanged, but $L_{\mathrm{s}}$ and
$\Lambda_{\mathrm{1.4}}$ are both substantially reduced, as shown by
Table \ref{TAB2}. In Fig.~\ref{fig:mr}, we present the $M$-$R$
curves determined by $\sigma\omega\rho2$ model with different values
of $L_{\mathrm{s}}$. Smaller values of $L_{\mathrm{s}}$ and
$\Lambda_{\mathrm{1.4}}$ correspond to smaller radii.

%\begin{widetext}

%\end{widetext}
\begin{widetext}

\begin{figure*}[htbp]
\centering
\includegraphics[width=5.8in]{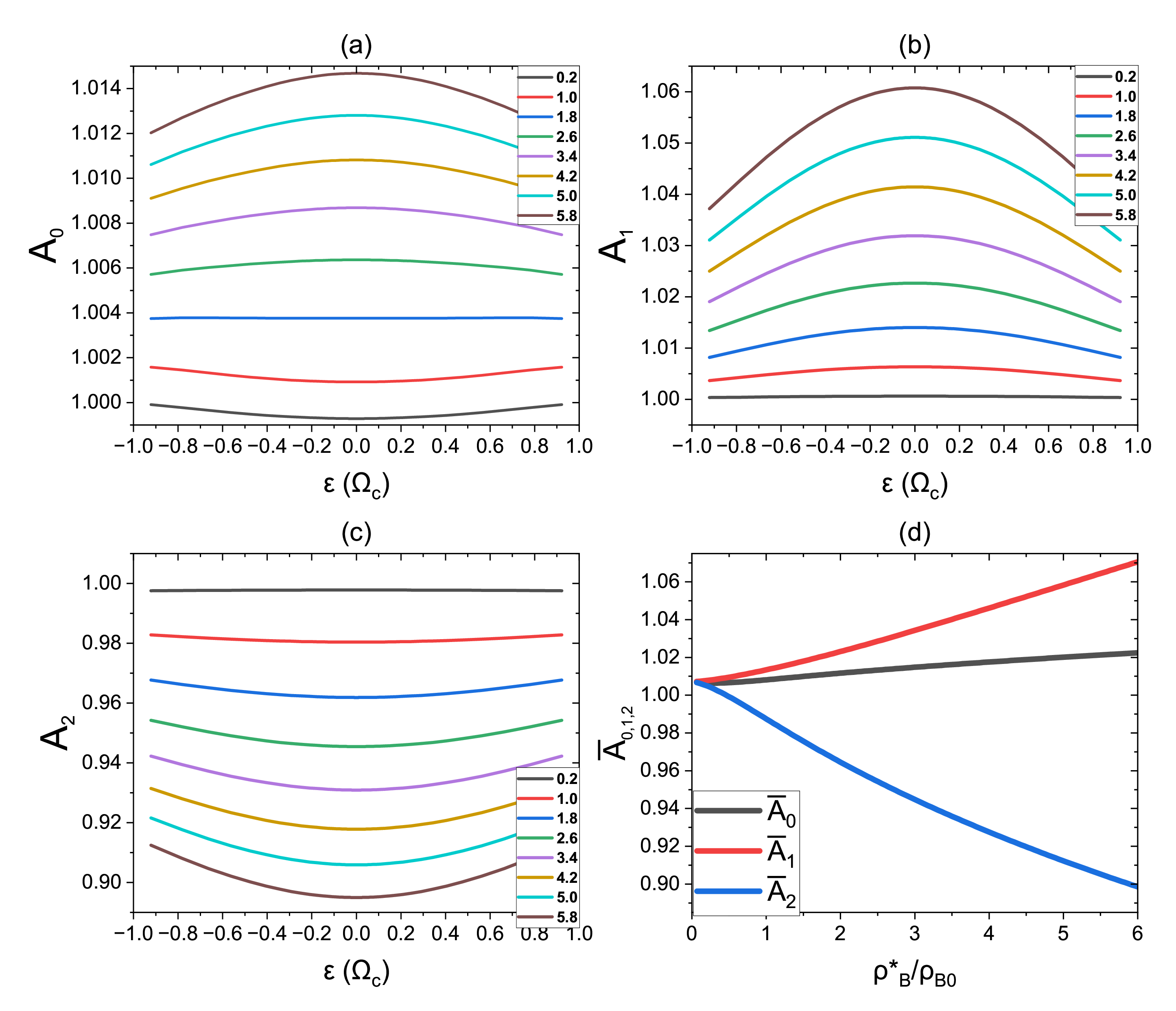}
\caption{Behavior of renormalization functions for
$\sigma\omega\rho2$ model at $L_{\mathrm{s}}=40~$MeV and
$a_{\rho}=0.7070$. (a - c) Energy dependence of $A_0(\varepsilon)$,
$A_1(\varepsilon)$, and $A_2(\varepsilon)$ for eight densities:
$0.2\rho_{\mathrm{B}0}$, $1.0 \rho_{\mathrm{B}0}$,
$1.8\rho_{\mathrm{B}0}$, $2.6\rho_{\mathrm{B}0}$,
$3.4\rho_{\mathrm{B}0}$, $4.2\rho_{\mathrm{B}0}$, $5.0
\rho_{\mathrm{B}0}$, $5.8\rho_{\mathrm{B}0}$. (d) Density dependence
of averaged quantities ${\bar A}_0$, ${\bar A}_1$, and ${\bar
A}_2$.} \label{fig:A012}
\end{figure*}

\end{widetext}

Shown in Fig.~\ref{fig:mr} are some recently observed values of
neutron star masses and radii. One can see that the three $M$-$R$
curves obtained with $L_{\mathrm{s}}=30~\mathrm{MeV}$,
$L_{\mathrm{s}}=40~\mathrm{MeV}$, and
$L_{\mathrm{s}}=50~\mathrm{MeV}$ are very consistent with the masses
and radii extracted from the GW170817 event generated by the binary
neutron star merger
($R_{1.4M_{\odot}}=11.85^{+1.95}_{-1.95}~\mathrm{km}$)
\cite{Annala18}, those from PSR J0030+0451
($M=1.44^{+0.15}_{-0.14}~M_{\odot}$ and
$R=13.02^{+1.24}_{-1.06}~\mathrm{km}$ reported by Miller \emph{et
al.} \cite{Miller19} and $M=1.34^{+0.15}_{-0.16}~M_{\odot}$ and
$R=12.71^{+1.14}_{-1.19}~\mathrm{km}$ reported by Riley \emph{et
al.} \cite{Riley19}), and those from PSR J0740+6620
($M=2.08^{+0.07}_{-0.07}~M_{\odot}$ and
$R=13.7^{+2.6}_{-1.5}~\mathrm{km}$ reported by Miller \emph{et al.}
\cite{Miller21} and $M=2.072^{+0.067}_{-0.066}~M_{\odot}$ and
$R=12.39^{+1.30}_{-0.98}~\mathrm{km}$ reported by Riley \emph{et
al.} \cite{Riley21}). These three $M$-$R$ curves also satisfy the
mass constraints of PSR J1614-2230
($1.97^{+0.04}_{-0.04}~M_{\odot}$) \cite{Demorest10}, PSR J0348+0432
($2.01^{+0.04}_{-0.04}~M_{\odot}$) \cite{Antoniadis13}, PSR
J0740+6620 ($2.14^{+0.10}_{-0.09}~M_{\odot}$) \cite{Cromartie19},
and PSR J0952-0607 ($2.35^{+0.17}_{-0.17}~M_{\odot}$)
\cite{Romani22}.

Recently, two binary compact objects were detected from the
gravitational wave event GW190814 \cite{Abbott20}. The heavier
object has a mass $\approx 23~M_{\odot}$ and was soon recognized as
a black hole \cite{Abbott20}. The other one has a much smaller mass
about $(2.50-2.67)$$M_{\odot}$. It remains an open puzzle whether
the lighter one is the heaviest known neutron star or the lightest
black hole ever found. The maximal mass predicted by the $L30$,
$L40$, $L50$ curves plotted in Fig.~\ref{fig:mr} is
$M_{\mathrm{max}} \approx 2.67M_{\odot}$. Thus, our theoretical
results point to the former possibility and indicate that such a
putative massive neutron star would have a radius within the range
of $(12.65-13.55)~$km. A more definite conclusion might be drawn by
measuring its radius with high precision.

For $L_{\mathrm{s}}=40~\mathrm{MeV}$, our theoretical result
$\Lambda_{1.4} = 872$ satisfies the constraint
$458\leq\Lambda_{\mathrm{1.4}} \leq 889$ inferred from GW190814 if
the low-mass compact object is assumed to be a neutron star
\cite{Abbott20}. In this view, the magnitudes of $M_{\mathrm{max}}$
and $\Lambda_{1.4}$ obtained in our calculations are consistent with
each other. However, the result $\Lambda_{1.4}= 872$ is out of the
range $70\leq\Lambda_{\mathrm{1.4}}\leq 580$ extracted from GW170817
\cite{Abbott17, Abbott18}. At present, the precise value of
$\Lambda_{1.4}$ remains largely uncertain due to the insufficiency
of observational data on the tidal deformability. Analysis based on
diverse assumptions and approximations often yield conflicting
results of $\Lambda_{1.4}$. More comprehensive gravitational wave
events from binary mergers are required to impose more stringent
constraints on $\Lambda_{1.4}$.

We plot the numerical solutions of $A_{0,1,2}(\varepsilon)$ in
Fig.~\ref{fig:A012} for symmetric nuclear matter at
$L_{\mathrm{s}}=40~$MeV and $a_{\rho}=0.7070$. Their properties in
neutron-dominated neutron star matter are similar. One can see from
Fig.~\ref{fig:A012}(d) that $\bar{A}_{0}$, $\bar{A}_{1}$, and
$\bar{A}_{2}$ are close but never equal to unity at very low
densities. With the increase of the density, $\bar{A}_{0}$,
$\bar{A}_{1}$, and $\bar{A}_{2}$ deviate considerably from unity.
Quantum many-body effects are more significant at higher densities.
However, they are present at all densities. We find that
$A_{0,1,2}(\varepsilon)$ have convergent solutions within the
density range $\rho^{\ast}_{\mathrm{B}} < 6\rho_{\mathrm{B}0}$,
which is much wider than the density range covered by $\chi$EFT.

\begin{figure}[htbp]
\centering
\includegraphics[width=3.8in]{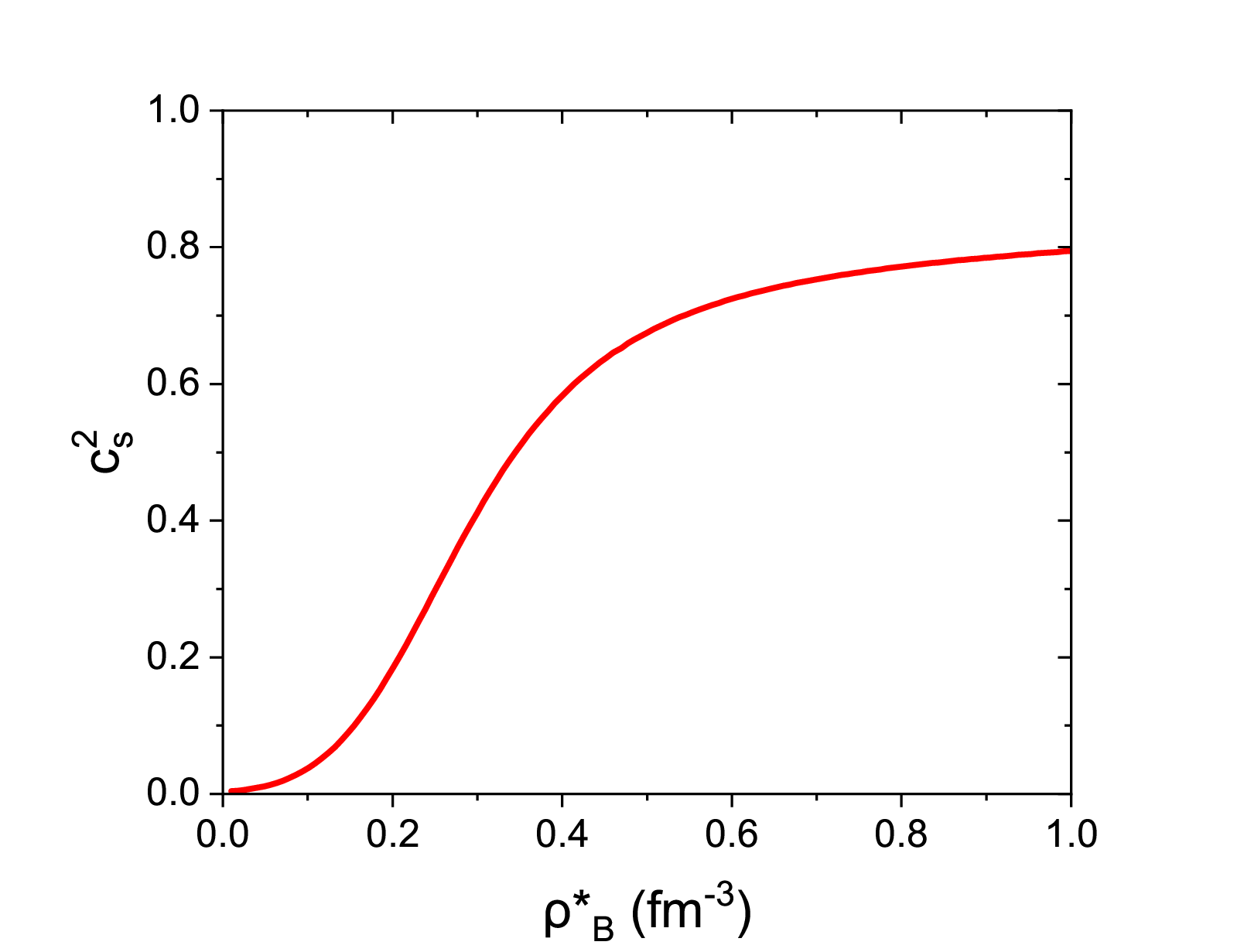}
\caption{The speed of sound squared as a function of neutron density
in a pure neutron matter. Here, we have made the speed of light
squared $c^2=1$.} \label{fig:soundspeed}
\end{figure}

The speed of sound $c_{s}$ is also a pivotal factor in the
description of neutron stars as it characterizes the stiffness and
incompressibility of the EOS \cite{Glendenningbook}. Causality
requires that $c_{s}$ must be smaller than the speed of light $c$.
We have computed the sound speed squared, defined by
\begin{eqnarray}
c_{s}^{2} = \frac{\partial P(\epsilon)}{\partial \epsilon},
\end{eqnarray}
for a pure neutron matter. The density dependence of $c_{s}^{2}$ is
depicted in Fig.~\ref{fig:soundspeed}. Evidently, the ratio
$c^{2}_{s}/c^{2}$ is always smaller than unity. As the neutron
density increases, the ratio converges towards a consistent value of
$0.8$. This conclusion also holds for symmetric nuclear matter and
neutron star matter. Therefore, the causality is ensured in all of
our results.

\section{Summary and discussion}\label{sec:summary}

In summary, we propose a nonperturbative approach to incorporate
the effects of quantum many-body correlations into the EOS of
neutron stars. The key step of this approach is to solve the
self-consistent DS equation of the renormalized neutron propagator
$G(k)$, which takes into account the many-body correlation effects
through three renormalization functions. After carrying out
extensive calculations, we select out a relatively simple
$\sigma$-$\omega$-$\rho$ model to describe the physics of dense
neutron star matter. We show that the EOS generated from this model
leads to several observable quantities of saturation nuclear matter
that are consistent with experiments. The $M$-$R$ relation
determined by such an EOS is comparable with some recent
astrophysical observations.

Our approach can be extended and further improved in several
directions. First, the zero-temperature EOS obtained in this paper
can be generalized to finite temperatures. Finite temperature EOS
may have applications in the description of protoneutron stars
\cite{Prakash97} and binary neutron star mergers \cite{Baiotti17}.
Second, our approach can be employed to examine the influence of
additional exotic degrees of freedom, such as hyperons
\cite{Glendenningbook, GM1} and $\Delta$-isobars
\cite{Li18, Sedrakian20}, that might be present in the high-density
region of neutron star core. These additional degrees of freedom may
soften the EOS and decrease the radii and/or the maximum mass
$M_{\mathrm{max}}$ of neutron stars. Third, the momentum dependence
of the current vertex functions $\Upsilon_{I}$,
$\Gamma_{\gamma_{\nu}}$, $\Theta_{\tau_3\gamma_{\nu}}$, and
$\Omega_{\tau_3}$ (see the last three paragraphs of the Appendix
\ref{sec:dsederivation} for a detailed explanation) need to be taken
into account. One possible way to do this is to expand these current
vertex functions around the Fermi momentum $k^{~}_{\mathrm{F}}$ by
taking $|\mathbf{k}|-k^{~}_{\mathrm{F}}$ as a small parameter.
Finally, the EOS obtained in this paper is calculated based on a
combination of the nonperturbative DS equation and RMFT. In an
effort to go beyond such a hybrid route, we would directly utilize
the renormalized neutron propagator $G(k)$, extracted from the
solutions of its DS equation, to compute thermodynamic quantities
and EOS by virtue of the LW functional \cite{Luttinger}.

As mentioned in Sec.~\ref{sec:eosofnss}, the tidal deformability
$\Lambda_{1.4}$ generated by our EOS appears to be considerably
larger than the result constrained by recent gravitational wave
events. In the future, we will examine whether a more realistic value
of $\Lambda_{1.4}$ could be obtained after some of the improvements
discussed above are accomplished.

The analysis in Sec.~\ref{sec:subseceos} and ~\ref{sec:eosofnss}
indicates that our approach offers a formalism different from RMFT.
It turns out that our results do not reduce to RMFT results at any
density. Given the impressive success of RMFT in the description of
finite nuclei, one might wonder whether our approach can be used to
study finite nuclei. Our non-perturbative formalism is rooted in the
quantum many-body theory, which is supposed to be applicable to an
equilibrium or near-equilibrium system that has a large number
($\approx 10^{23}$) of particles. The neutron-star matter can be
approximately considered as such a system. The lower limit
$|\mathbf{q}|=0$ for meson momentum $\mathbf{q}$ implies that the
neutron-star matter is assumed to have a very large volume. However,
finite nuclei means a small system containing typically dozens of
nucleons. The nonperturbative formalism cannot be directly employed
to deal with finite nuclei. In order to embody the smallness of
finite nuclei volume, the meson momentum $|\mathbf{q}|$ should be
larger than some critical value $\Lambda_{\mathrm{IR}}$, which
serves as an infrared cutoff. Accordingly, the integration range
$|\mathbf{q}|\in [0,\Lambda_{\mathrm{c}} k^{~}_{\mathrm{F}}]$ should
be replaced by $|\mathbf{q}|\in [\Lambda_{\mathrm{IR}},
\Lambda_{\mathrm{c}} k^{~}_{\mathrm{F}}]$. Then one could solve the
integral equations of $A_{0,1,2}(\varepsilon)$ and utilize the
solutions to describe the properties of finite nuclei. Our present
work concentrates on the many-body correlation effects in
neutron-star matter. It would be of interest to examine whether the
properties of finite nuclei can be described by a reformed version
of our formalism.

We finally remark on the application of our approach to the
investigation of neutron superfluidity. It is universally believed
\cite{Sedrakian, Page} that neutron superfluid exists in neutron
stars and is responsible for the observed glitch phenomena.
Superfluid also plays a significant role in the thermal evolution of
neutron stars \cite{Page}. Unfortunately, both the superfluid gap
$\Delta_{n}$ and transition temperature $T_{c}$ have not been
unambiguously determined to date. Previous calculations of
$\Delta_{n}$ and $T_{c}$ are predominantly based on BCS mean-field
theory \cite{Sedrakian, Page}. The pairing potential $V(\mathbf{r})$
used in such calculations is obtained from the experimental data of
two-body scattering phase shift \cite{Sedrakian, Page}. These
mean-field calculations are problematic for several reasons. The
first one is that the instantaneous pairing potential entirely
neglects the retardation of meson propagation. The second one is the
absence of the in-medium effect. Moreover, BCS mean-field theory fails
to include the influence of neutron damping and neutron
velocity/mass renormalization. Another obvious shortcoming of
BCS-level calculations is that the pairing potential $V(\mathbf{r})$
is entirely unknown at high densities due to the absence of
experimental data of scattering phase shift above $350~\mathrm{MeV}$.
For a massive neutron star, the relevant neutron energy in the
high-density region could be much greater than $350~\mathrm{MeV}$.
For these reasons, it is important to study neutron superfluidity
beyond BCS mean-field theory. The nonperturbative approach that we
develop in this paper is capable of including the effects ignored in
previous BCS-level calculations, such as the neutron damping, the
neutron velocity/mass renormalization, the retardation of meson
propagation, as well as the in-medium screening (it can be embodied
by adding the polarization functions to the meson propagators).
Furthermore, our DS equation has as convergent solutions even when the
neutron density is as high as $6\rho_{\mathrm{B}}$, which is far
beyond $350~\mathrm{MeV}$. We thus expect that our approach provides
a more efficient framework for the theoretical study of neutron
superfluidity than the BCS mean-field theory. This issue will be
addressed in a separate work.

Some alternative forms of exotic matter, such as quark stars and hybrid
stars, have been proposed to understand the physics of neutron stars
\cite{Glendenningbook}. Especially, quark degrees of freedom may
exist in the cores of certain neutron stars \cite{Annala20,
Annala23}. The interactions of quarks with mesons or gluons might
result in significant quantum many-body effects. It would be
interesting to study such effects by employing an extended version
of the nonperturbative DS equation framework.

\section{Acknowledgement}

We thank Sophia Han, Wei-Zhou Jiang, and Ang Li for helpful
discussions. This work is supported by the National Natural Science
Foundation of China under Grant No. 12073026, the Anhui Natural
Science Foundation under Grant No. 2208085MA11, and the
Fundamental Research Funds for the Central Universities.

\appendix

\section{Derivation of Dyson-Schwinger equation of nucleon propagator
\label{sec:dsederivation}}

Here we provide a detailed derivation of the DS equation of the
neutron propagator given by Eq.~(\ref{eq:dselwmodel}). The
derivation will be performed within the path-integral framework of
quantum field theory. The Lagrangian density Eq.~(\ref{eq:LWmodel})
can be re-written as
\begin{eqnarray}
\mathcal{L}_{\mathrm{T}} &=& \bar{\psi}(i\partial_{\mu}\gamma^{\mu}
- m^{~}_{\mathrm{N}})\psi+\frac{1}{2}\partial_{\mu} \sigma
\partial^{\mu}\sigma-\frac{1}{2} m^{2}_{\sigma}\sigma^{2}
\nonumber \\
&& -\frac{1}{4}\omega_{\mu\nu}\omega^{\mu\nu}
+ \frac{1}{2}m^{2}_{\omega}\omega_\mu\omega^{\mu} \nonumber \\
&& -\frac{1}{4}\rho_{3\mu\nu}\rho_3^{\mu\nu} +
\frac{1}{2}m^{2}_{\rho} \rho_{3\mu}\rho_3^{\mu}\nonumber\\
&& +\frac{1}{2}\partial_{\mu}\delta_3\partial^\mu
\delta_{3} -\frac{1}{2}m^{2}_{\delta}\delta_{3}^{2} \nonumber\\
&& +g_{\sigma}\sigma \bar{\psi}\psi-g_{\omega}\omega_{\mu}
\bar{\psi}\gamma^{\mu}\psi\nonumber\\
&& -\frac{1}{2}g_{\rho}\rho_{3\mu}\bar{\psi}\tau_3\gamma^{\mu} \psi
+ g_\delta\delta_{3}\bar{\psi}\tau_{3}\psi \nonumber \\
&& + \bar{\eta} {\psi}+\bar{\psi} \eta +
J_{\sigma}\sigma+J^{\mu}_{\omega}\omega_{\mu} \nonumber \\
&& +J^{\mu}_{\rho} \rho_{3\mu}+J_{\delta}\delta_{3}.
\label{applagrangiandensity}
\end{eqnarray}
Here, we have introduced several external sources $J_{\sigma}$,
$J^{\mu}_{\omega}$, $J^{\mu}_{\rho}$, $J_{\delta}$, $\bar{\eta}$,
and $\eta$, which are associated with fields $\sigma$,
$\omega_{\mu}$, $\rho_{3\mu}$, $\delta_{3}$, ${\psi}$, and
$\bar{\psi}$, respectively.

To help readers understand the calculational details, we list some
basic rules of functional integral \cite{Itzykson, Liu21}. All the
correlation functions are generated from three quantities: the
partition function $Z\left[J_{\sigma},J^{\mu}_{\omega},
J^{\mu}_{\rho},J_{\delta},\bar{\eta},\eta\right]$, the generating
functional $W\left[J_{\sigma},J^{\mu}_{\omega},
J^{\mu}_{\rho},J_{\delta},\bar{\eta},\eta\right]$ and the generating
functional $\Xi\left[\sigma,\omega_{\mu},\rho_{3\mu},\delta_{3},
{\psi},\bar{\psi}\right]$. They are defined as:
\begin{eqnarray}
Z\left[J_{\sigma},J^{\mu}_{\omega},J^{\mu}_{\rho},J_{\delta},\bar{\eta},
\eta\right] &=& \int\mathcal{D}\sigma \mathcal{D}\omega_{\mu}
\mathcal{D}\rho_{3\mu}\mathcal{D}\delta_{3}\mathcal{D}\bar{\psi}
\mathcal{D}\psi \nonumber \\
&& \times \exp{\left(i\int d^{4}z\mathcal{L}_\mathrm{T}\right)},
\label{appz} \nonumber\\
W\left[J_{\sigma},J^{\mu}_{\omega},J^{\mu}_{\rho},J_{\delta},\bar{\eta},
\eta\right] &=& -i\ln Z\left[J_{\sigma},J^{\mu}_{\omega},
J^{\mu}_{\rho},J_{\delta},\bar{\eta},\eta\right], \label{appw} \nonumber \\
\Xi\left[\sigma,\omega_{\mu},\rho_{3\mu},\delta_{3},{\psi},\bar{\psi}\right]
&=& W\left[J_{\sigma},J^{\mu}_{\omega}, J^{\mu}_{\rho},
J_{\delta},\bar{\eta}, \eta\right] \nonumber \\
&& -\int\big(J_{\sigma}\sigma+J^{\mu}_{\omega}\omega_{\mu} \nonumber \\
&& +J^{\mu}_{\rho}\rho_{3\mu}+J_{\delta}\delta_{3}+\bar{\eta}
{\psi}+\bar{\psi}\eta\big).\label{appxi}\nonumber
\end{eqnarray}
Here, $z=(t,\bm{z})$ is $(1+3)$-dimensional position vector. The
following identities will be frequently used:
\begin{eqnarray}
&&\frac{\delta W}{\delta J_{\sigma}}=\langle\sigma\rangle, \qquad
\frac{\delta W}{\delta J^{\mu}_{\omega}}=\langle\omega_{\mu}\rangle,
\qquad \frac{\delta W}{\delta J^{\mu}_{\rho}} =
\langle\rho_{3\mu}\rangle, \nonumber \\
&& \frac{\delta W}{\delta J_{\delta}}=\langle\delta_{3}\rangle,
\qquad \frac{\delta W}{\delta \eta} = -\langle\bar{\psi}\rangle,
\qquad \frac{\delta W}{\delta \bar{\eta}}=\langle\psi\rangle,
\label{eq:functionalrules1}\nonumber \\
&& \frac{\delta \Xi}{\delta \sigma}=- J_{\sigma}, \qquad
\frac{\delta \Xi}{\delta \omega_{\mu}}=- J^{\mu}_{\omega}, \qquad
\frac{\delta \Xi}{\delta \rho_{3\mu}}=- J^{\mu}_{\rho}, \nonumber \\
&& \frac{\delta \Xi}{\delta \delta_{3}}=- J_{\delta},
\qquad\frac{\delta \Xi}{\delta\psi}= \bar{\eta}, \qquad\frac{\delta
\Xi}{\delta \bar{\psi}}=-\eta.\nonumber
\label{eq:functionalrules2}
\end{eqnarray}

Generating functional $W$ generates various connected correlation
functions, whereas $\Xi$ generates all the irreducible proper
vertices. In Lagrangian density $\mathcal{L}_{\mathrm{T}}$ there are
four different interactions between mesons and nucleons, denoted by
$\sigma N$, $\omega N$, $\rho_0 N$, and $\delta_0 N$, respectively.
We presume the system exhibits translational invariance.
The full (exact) nucleon propagator $G(z-z_1)$, the full
$\sigma$-meson propagator $D(z-z_2)$, the full $\omega$-meson
propagator $F^{\mu\nu}(z-z_2)$, the full $\rho_0$-meson propagator
$V^{\mu\nu}(z-z_2)$, and the full $\delta_0$-meson propagator
$C(z-z_2)$ are defined in terms of $W$ in order by
\begin{eqnarray}
G(z-z_1) &\equiv& -i\langle\psi(z)\bar{\psi}(z_1)\rangle =
\frac{\delta^{2}W}{\delta\bar{\eta}(z)\delta\eta(z_1)},
\label{nucleonpropagator}\nonumber \\
D(z-z_2) &\equiv& -i\langle\sigma(z)\sigma^\dag(z_2)\rangle =
-\frac{\delta^{2}W}{\delta J_{\sigma}(z)\delta J_{\sigma}(z_2)},
\label{sigmapropagator}\nonumber \\
F^{\mu\nu}(z-z_2) &\equiv& -i\langle\omega^{\mu}(z)
\omega^{\nu\dag}(z_2)\rangle = -\frac{\delta^{2}W}{\delta
J^{\mu}_{\omega}(z)\delta J^{\nu}_{\omega}(z_2)},
\label{omegapropagator}\nonumber \\
V^{\mu\nu}(z-z_2) &\equiv& -i\langle\rho_3^{\mu}(z)
\rho_3^{\nu\dag}(z_2)\rangle = -\frac{\delta^{2}W}{\delta
J^{\mu}_{\rho}(z)\delta J^{\nu}_{\rho}(z_2)},
\label{rhopropagator}\nonumber \\
C(z-z_2) &\equiv& -i\langle\delta_3^{\mu}(z)
\delta_3^{\dag}(z_2)\rangle = -\frac{\delta^{2}W}{\delta
J_{\delta}(z)\delta J_{\delta}(z_2)}.\nonumber
\label{deltapropagator}
\end{eqnarray}
The vertex corrections to nucleon-meson interactions can be
described by four three-point correlation functions $\langle \sigma
\psi\bar{\psi}\rangle$, $\langle \omega^{\mu}
\psi\bar{\psi}\rangle$, $\langle \rho_3^{\mu}
\psi\bar{\psi}\rangle$, and $\langle \delta_3
\psi\bar{\psi}\rangle$, which are defined via $W$ and $\Xi$ as
follows:
\begin{eqnarray}
\langle\sigma\psi\bar{\psi}\rangle &\equiv& \frac{\delta^3W}{\delta
J_{\sigma}\delta\bar{\eta} \delta\eta} = -DG\frac{\delta^3
\Xi}{\delta \sigma\delta\bar{\psi}\delta\psi}G, \label{eq:3pcfsigma} \\
\langle\omega^{\mu}\psi\bar{\psi}\rangle &\equiv&
\frac{\delta^3W}{\delta J^{\mu}_{\omega}\delta\bar{\eta} \delta\eta}
= -F_{\mu\nu}G\frac{\delta^3\Xi}{\delta \omega_{\nu}\delta\bar{\psi}
\delta\psi}G, \label{eq:3pcfomega} \\
\langle\rho_3^{\mu}\psi\bar{\psi}\rangle &\equiv&
\frac{\delta^3W}{\delta J^{\mu}_{\rho}\delta\bar{\eta} \delta\eta} =
-V_{\mu\nu}G\frac{\delta^3\Xi}{\delta \rho_{3\nu}\delta\bar{\psi}
\delta\psi}G,\label{eq:3pcfrho}\\
\langle\delta_3\psi\bar{\psi}\rangle &\equiv&
\frac{\delta^3W}{\delta J_{\delta}\delta\bar{\eta} \delta\eta} =
-CG\frac{\delta^3\Xi}{\delta \delta_{3}\delta\bar{\psi}
\delta\psi}G. \label{eq:3pcfdelta}
\end{eqnarray}

Next, we derive the DS equation of the full nucleon propagator with
the help of the above identities. Given that $Z$ is invariant under
an arbitrary infinitesimal variation of spinor field $\bar{\psi}$,
i.e.,
\begin{eqnarray}
\int \mathcal{D}\sigma \mathcal{D}\omega_{\mu}\mathcal{D}
\rho_{3\mu}\mathcal{D}\delta_{3}\mathcal{D} \bar{\psi}
\mathcal{D}{\psi}\frac{\delta}{i\delta\bar{\psi}}\exp{\left(i\int
d^4z\mathcal{L}_\mathrm{T}\right)} = 0,\nonumber
\label{eq:Partition}
\end{eqnarray}
we find that the following relation holds:
\begin{eqnarray}
&&\langle(i\partial_{\mu}\gamma^{\mu}-m_{\mathrm{N}})\psi(z)\rangle
+ \langle \eta(z)\rangle \nonumber \\
&& +\langle g_{\sigma}\sigma(z) \psi(z)\rangle - \langle
g_{\omega}\omega_{\mu}(z)\gamma^{\mu}\psi(z)\rangle \nonumber \\
&& -\langle \frac{g_{\rho}}{2}\rho_{3\mu}(z)\tau_3
\gamma^{\mu}\psi(z)\rangle + \langle g_{\delta} \delta_{3}(z)
\tau_3\psi(z)\rangle = 0.\nonumber \label{eq:eomfermion}
\end{eqnarray}
This relation is equivalent to
\begin{widetext}
\begin{eqnarray}
-\eta(z) &=& (i\partial_{\mu}\gamma^{\mu} - m_{\mathrm{N}})
\frac{\delta W}{\delta \bar{\eta}(z)} - ig_{\sigma}\frac{\delta^{2}
W}{\delta J_{\sigma}(z) \delta \bar{\eta}(z)} +
g_{\sigma}\frac{\delta W}{\delta J_{\sigma}(z)} \frac{\delta
W}{\delta \bar{\eta}(z)}+ig_{\omega}\gamma^{\mu}\frac{\delta^{2}
W}{\delta J^{\mu}_{\omega}(z)\delta
\bar{\eta}(z)}-g_{\omega}\frac{\delta W}{\delta
J^{\mu}_{\omega}(z)}\gamma^{\mu}\frac{\delta W}{\delta
\bar{\eta}(z)} \nonumber \\
&& +i\frac{g_{\rho}}{2}\tau_3\gamma^{\mu}\frac{\delta^{2} W}{\delta
J^{\mu}_{\rho}(z)\delta\bar{\eta}(z)}-\frac{g_{\rho}}{2}
\frac{\delta W}{\delta J^{\mu}_{\rho}(z)}\tau_3\gamma^{\mu}
\frac{\delta W}{\delta \bar{\eta}(z)} - ig_{\delta}\tau_3
\frac{\delta^{2} W}{\delta J_{\delta}(z)\delta\bar{\eta}(z)} +
g_{\delta} \frac{\delta W}{\delta J_{\delta}(z)}\tau_3 \frac{\delta
W}{\delta \bar{\eta}(z)}.\label{eq:eomfermion2}
\end{eqnarray}
Two terms of the above equation vanish upon removing sources and can
be directly omitted. Carrying out functional derivatives of both
sides of this equation with respect to $\eta(z_2)$ and using the
identities (\ref{eq:3pcfsigma}-\ref{eq:3pcfdelta}) leads to
\begin{eqnarray}
\delta(z-z_2) &=& (i\partial_{\mu}\gamma^{\mu}-m_{\mathrm{N}})
\frac{\delta^{2} W}{\delta \bar{\eta}(z)\delta\eta(z_2)} -
ig_{\sigma}\frac{\delta^3 W}{\delta J_{\sigma}(z)\delta
\bar{\eta}(z)\delta \eta(z_2)} + ig_{\omega}\gamma^{\mu}
\frac{\delta^3 W}{\delta J^{\mu}_{\omega}(z) \delta
\bar{\eta}(z)\delta \eta(z_2)}
\nonumber \\
&& +i\frac{g_{\rho}}{2}\tau_3\gamma^{\mu}\frac{\delta^3 W}{\delta
J^{\mu}_{\rho}(z)\delta\bar{\eta}(z)\delta \eta(z_2)} -
ig_{\delta}\tau_3\frac{\delta^3 W}{\delta J_{\delta}(z)
\delta\bar{\eta}(z)\delta \eta(z_2)}  \nonumber \\
&=&(i\partial_{\mu}\gamma^{\mu}-m_{\mathrm{N}})G(z-z_2) +
ig_{\sigma}\int dz_1 dz_3 dz_4D(z-z_3)G(z-z_1)\frac{\delta^3
\Xi}{\delta\sigma(z_3) \delta\bar{\psi}(z_1)\delta {\psi}(z_4)}
G(z_4-z_2) \nonumber \\
&& -ig_{\omega}\int dz_1 dz_3 dz_4 \gamma_{\mu}F^{\mu\nu}(z-z_3)
G(z-z_1)\frac{\delta^{3}\Xi}{\delta \omega^{\nu}(z_3)
\delta\bar{\psi}(z_1)\delta {\psi}(z_4)}G(z_4-z_2) \nonumber \\
&& -i\frac{g_{\rho}}{2}\int dz_1 dz_3 dz_4 \tau_3\gamma_{\mu}
V^{\mu\nu}(z-z_3)G(z-z_1)\frac{\delta^3 \Xi}{\delta
\rho^{\nu}_{3}(z_3) \delta\bar{\psi}(z_1)\delta
{\psi}(z_4)}G(z_4-z_2)\nonumber \\
&& +ig_{\delta}\int dz_1 dz_3 dz_4 \tau_{3}C(z-z_3)G(z-z_1)
\frac{\delta^3 \Xi}{\delta \delta_{3}(z_3) \delta\bar{\psi}(z_1)
\delta {\psi}(z_4)}G(z_4-z_2).\nonumber \label{eq:eomfermion3}
\end{eqnarray}
\end{widetext}
This equation can be further expressed as
\begin{eqnarray}
&& G^{-1}(z-z_2) \nonumber \\
&=& (i\partial_{\mu}\gamma^{\mu} - m_{\mathrm{N}})\delta(z-z_2)
\nonumber \\
&& +ig_{\sigma}\int dz_1 dz_3 D(z-z_3)G(z-z_1)\Upsilon(z_3,z_1,z_2)
\nonumber \\
&& -ig_{\omega}\int dz_1 dz_3 \gamma_{\mu}F^{\mu\nu}(z-z_3)G(z-z_1)
\Gamma_{\nu}(z_3,z_1,z_2)\nonumber \\
&& -i\frac{g_{\rho}}{2}\int dz_1 dz_3 \tau_3\gamma_{\mu}
V^{\mu\nu}(z-z_3)G(z-z_1)\Theta_{\nu}(z_3,z_1,z_2)\nonumber\\
&& +ig_{\delta}\int dz_1 dz_3 \tau_3 C(z-z_3)G(z-z_1)
\Omega(z_3,z_1,z_2).\label{eq:eomfermion4}
\end{eqnarray}
Here, we have defined a truncated (all external legs being dropped)
$\sigma N$ interaction vertex function
\begin{eqnarray}
\Upsilon(z_3,z_1,z_2) = \frac{\delta^3 \Xi}{\delta \sigma(z_3)
\delta \bar{\psi}(z_1)\delta {\psi}(z_2)}, \label{eq:upsilon}
\end{eqnarray}
a truncated $\omega N$ interaction vertex function
\begin{eqnarray}
\Gamma_{\nu}(z_3,z_1,z_2) = \frac{\delta^3 \Xi}{\delta
\omega^{\nu}(z_3) \delta \bar{\psi}(z_1)\delta {\psi}(z_2)},
\label{eq:gammamu}
\end{eqnarray}
a truncated $\rho_0 N$ interaction vertex function
\begin{eqnarray}
\Theta_{\nu}(z_3,z_1,z_2) = \frac{\delta^3 \Xi}{\delta
\rho^\nu_{3}(z_3) \delta \bar{\psi}(z_1)\delta{\psi}(z_2)},
\label{eq:thetanu}
\end{eqnarray}
and a truncated $\delta_0 N$ interaction vertex function
\begin{eqnarray}
\Omega(z_3,z_1,z_2) = \frac{\delta^3 \Xi}{\delta \delta_{3}(z_3)
\delta \bar{\psi}(z_1)\delta{\psi}(z_2)}. \label{eq:omega}
\end{eqnarray}
The propagators are Fourier transformed as
\begin{eqnarray}
G(z-z_1) &=& \int\frac{d^4k}{(2\pi)^4}e^{-ik(z-z_1)}
G(k),\label{eq:gfourier} \nonumber \\
D(z-z_2) &=& \int\frac{d^4q}{(2\pi)^4}e^{-iq(z-z_2)}D(q),
\label{eq:dfourier} \nonumber \\
F^{\mu\nu}(z-z_2) &=& \int\frac{d^4q}{(2\pi)^4}
e^{-iq(z-z_2)}F^{\mu\nu}(q),\label{eq:ffourier} \nonumber \\
V^{\mu\nu}(z-z_2) &=& \int\frac{d^4q}{(2\pi)^4}
e^{-iq(z-z_2)}V^{\mu\nu}(q),\label{eq:vfourier} \nonumber \\
C(z-z_2) &=& \int\frac{d^4q}{(2\pi)^4}e^{-iq(z-z_2)}C(q).
\label{eq:cfourier} \nonumber
\end{eqnarray}
If the translational invariance is preserved, interaction vertex
functions are Fourier transformed as
\begin{eqnarray}
&& \Upsilon(z_{1},z,z_{2}) \equiv \Upsilon(z_{1}-z,z-z_{2})\nonumber \\
&=& \int\frac{d^4q d^4k}{(2\pi)^8}e^{-i(k+q)(z_1-z)}e^{-ik(z-z_2)}
\Upsilon(q,k), \label{eq:upsilonfourier} \nonumber \\
&& \Gamma_{\nu}(z_{1},z,z_{2}) \equiv \Gamma_{\nu}(z_1-z, z-z_2)\nonumber \\
&=& \int\frac{d^4q d^4k}{(2\pi)^8}e^{-i(k+q)(z_1-z)}e^{-ik(z-z_2)}
\Gamma_{\nu}(q,k),\label{eq:gammamufourier} \nonumber \\
&& \Theta_{\nu}(z_{1},z,z_{2}) \equiv \Theta_{\nu}(z_1-z,z-z_2)
\nonumber \\
&=& \int\frac{d^4q d^4k}{(2\pi)^{8}}e^{-i(k+q)(z_1-z)}e^{-ik(z-z_2)}
\Theta_{\nu}(q,k),
\label{eq:thetanufourier} \nonumber \\
&& \Omega(z_{1},z,z_{2}) \equiv \Omega(z_1-z, z-z_2)\nonumber \\
&=& \int\frac{d^4q d^4k}{(2\pi)^8} e^{-i(k+q)(z_1-z)}
e^{-ik(z-z_2)}\Omega(q,k).\nonumber \label{eq:omegafourier}
\end{eqnarray}
Performing Fourier transformation of Eq.~(\ref{eq:eomfermion4}) leads
us to the DS equation obeyed by the nucleon propagator:
\begin{eqnarray}
G^{-1}(k)&=& G_0^{-1}(k)+ig_{\sigma} \int \frac{d^4q}{(2\pi)^4}
G(k+q)D(q)\Upsilon(q,k) \nonumber \\
&& -ig_{\omega}\int \frac{d^4 q}{(2\pi)^4}\gamma_{\mu}
G(k+q)F^{\mu\nu}(q)\Gamma_{\nu}(q,k) \nonumber \\
&&-i\frac{g_{\rho}}{2}\int \frac{d^4q}{(2\pi)^4}\tau_3\gamma_{\mu}
G(k+q)V^{\mu\nu}(q)\Theta_{\nu}(q,k) \nonumber \\
&&+ig_{\delta}\int \frac{d^4q}{(2\pi)^4}\tau_3
G(k+q)C(q)\Omega(q,k). \label{eq:originaldseg}
\end{eqnarray}
Here, the free nucleon propagator has the form
\begin{eqnarray}
G_{0}(k)=\frac{1}{k_{\mu} \gamma^{\mu}-m^{~}_{\mathrm{N}}}.
\end{eqnarray}
The free propagators of $\sigma$, $\omega$, $\rho_0$, and $\delta_0$
are
\begin{eqnarray}
D_0(q) &=& \frac{1}{q^{2}-m^{2}_{\sigma}}, \\
F^{\mu\nu}_0(q) &=& \frac{-1}{{q}^{2} - m^{2}_{\omega}}
\left(g^{\mu\nu} - \frac{{q}^{\mu}{q}^{\nu}}{m^{2}_{\omega}}\right),\\
V^{\mu\nu}_0(q) &=& \frac{-1}{{q}^{2} - m^{2}_{\rho}}
\left(g^{\mu\nu} - \frac{{q}^{\mu}{q}^{\nu}}{m^{2}_{\rho}}\right),\\
C_0(q) &=& \frac{1}{q^{2}-m^{2}_{\delta}}.
\end{eqnarray}

The DS equations of meson propagators can be derived in an analogous
way \cite{Itzykson, Liu21}. We will not give the derivational
details and just present their final expressions:
\begin{widetext}
\begin{eqnarray}
D^{-1}(q) &=& D_{0}^{-1}(q)-ig_{\sigma}\int \frac{d^4k}{(2\pi)^4}
\mathrm{Tr}[G(k+q)\Upsilon(q,k)G(k)], \label{eq:dsed} \nonumber \\
F^{-1}_{\mu\nu}(q) &=& F^{-1}_{0\mu\nu}(q)+ ig_{\omega}\int
\frac{d^4k}{(2\pi)^4} \mathrm{Tr}[\gamma^{\mu}G(k+q)
\Gamma^{\nu}(q,k)G(k)],
\label{eq:dsefq} \nonumber \\
V^{-1}_{\mu\nu}(q) &=& V^{-1}_{0\mu\nu}(q)+i\frac{g_{\rho}}{2}
\int\frac{d^4k}{(2\pi)^4}\mathrm{Tr}[\tau_3\gamma^{\mu}
G(k+q)\Theta^{\nu}(q,k)
G(k)],\label{eq:dsevq}\nonumber \\
C^{-1}(q) &=& C^{-1}_{0}(q)-ig_{\delta}\int\frac{d^4k}{(2 \pi)^4}
\mathrm{Tr}[\tau_{3}G(k+q)\Omega(q,k)G(k)].\nonumber
\label{eq:dsevq}
\end{eqnarray}
The vertex functions $\Upsilon(q,k)$, $\Gamma_{\nu}(q,k)$,
$\Theta_{\nu}(q,k)$, and $\Omega(q,k)$ satisfy their own DS
equations, which are related to an infinite number of multipoint
correlation functions through an infinite number of DS equations.
The complete set of DS equations are exact, but are apparently too
complicated to handle.

The mutually coupled DS equations can be simplified by using a
method proposed in Ref.~\cite{Liu21}. The essential operation of
this method is to replace the product of a full meson propagator,
say $D(q)$, and the related interaction vertex function by the
product of the corresponding free meson propagator $D_{0}(q)$ and a
special current vertex functions. To explain how this method works,
we define some composite operators:
\begin{eqnarray}
J_{I}(z) &=&\bar{\psi}(z)I_{4\times4}\psi(z), \\
J^{\nu}(z) &=& \bar{\psi}(z)\gamma^{\nu}\psi(z), \\
J_3^{\nu}(z) &=& \bar{\psi}(z)\tau_3\gamma^{\nu}\psi(z),\\
J_3(z) &=& \bar{\psi}(z)\tau_3\psi(z),\label{eq:currentoperators}
\end{eqnarray}
where $\nu=0,1,2,3$. These composite operators are called current
operators because their forms are similar to various (e.g., scalar,
vector, isospin vector, and isospin scalar) currents. These current
operators are then used to define a number of current vertex
functions:
\begin{eqnarray}
\langle J_{I}(z)\psi(z_1)\bar{\psi}(z_2)\rangle &=& \langle
\bar{\psi}(z)I_{4\times4}\psi(z)\psi(z_1)\bar{\psi}(z_2)\rangle =
-\int dz_3 dz_4 G(z_1-z_3)\Upsilon_{I}(z,z_3,z_4)G(z_4-z_2),
\label{eq:currentvertexJI}\\
\langle J^{\nu}(z)\psi(z_1)\bar{\psi}(z_2)\rangle &=& \langle
\bar{\psi}(z)\gamma^{\nu}\psi(z)\psi(z_1)\bar{\psi}(z_2)\rangle =
-\int dz_3 dz_4 G(z_1-z_3)\Gamma_{\gamma^{\nu}}(z,z_3,z_4)
G(z_4-z_2),
\label{eq:currentvertexJmu} \\
\langle J_3^{\nu}(z)\psi(z_1)\bar{\psi}(z_2)\rangle &=& \langle
\bar{\psi}(z)\tau_3\gamma^{\nu}\psi(z)
\psi(z_1)\bar{\psi}(z_2)\rangle = -\int dz_3 dz_4
G(z_1-z_3)\Theta_{\tau_3\gamma^{\nu}}(z,z_3,z_4)G(z_4-z_2),
\label{eq:currentvertexJ3mu} \\
\langle J_3(z)\psi(z_1)\bar{\psi}(z_2)\rangle &=& \langle
\bar{\psi}(z)\tau_3\psi(z)\psi(z_1)\bar{\psi}(z_2)\rangle = -\int
dz_3 dz_4 G(z_1-z_3)\Omega_{\tau_3}(z,z_3,z_4)G(z_4-z_2).
\label{eq:currentvertexJ3}
\end{eqnarray}
\end{widetext}
Here, $\Upsilon_{I}$, $\Gamma_{\gamma^{\nu}}$,
$\Theta_{\tau_3\gamma^{\nu}}$, and $\Omega_{\tau_3}$ are called
current vertex functions. Now we take the $\omega N$ coupling as an
example to demonstrate how to derive the relation satisfied by
interaction and current vertex functions \cite{Liu21}. Making use of
the invariance of $Z$ under an infinitesimal variation of
$\omega$-meson field $\omega_{\mu}$, we derive an identity
\begin{eqnarray}
&& g_{\omega}\langle \bar{\psi}(z)\gamma^{\mu} \psi(z)\rangle
\nonumber \\
&=&\left[(\partial^{2}+m_{\omega}^{2})g^{\mu\nu}-\partial^{\mu}
\partial^{\nu}\right]\omega_{\nu}(z) + J^{\mu}_{\omega}(z),
\label{eq:VERTEX1}\nonumber
\end{eqnarray}
which is then cast into an equivalent form
\begin{eqnarray}
&& g_{\omega}\langle \bar{\psi}(z)\gamma^{\mu} \psi(z)\rangle
\nonumber \\
&=&\left[(\partial^{2}+m_{\omega}^{2})g^{\mu\nu}-\partial^{\mu}
\partial^{\nu}\right]\frac{\delta W}{\delta J^{\nu}_{\omega}(z)}
+J^{\mu}_{\omega}(z).\label{eq:VERTEX2}\nonumber
\end{eqnarray}
Carrying out functional derivatives with respect to $\bar{\eta}$ and
$\eta$ in order leads to
\begin{eqnarray}
&& \frac{\delta^{2}}{\delta\bar{\eta}(z_1)\delta\eta(z_2)}\langle
\bar{\psi}(z)\gamma^{\mu}\psi(z)\rangle \nonumber \\
&=& \langle \bar{\psi}(z)\gamma^{\mu}\psi(z){\psi}(z_1)
\bar{\psi}(z_2)\rangle \nonumber \\
&=& g_{\omega}^{-1}\Big[(\partial^{2}+m_{\omega}^{2})g^{\mu\nu} -
\partial^{\mu}\partial^{\nu}\Big]\frac{\delta^3 W}{\delta
J^{\nu}_{\omega}(z)\delta\bar{\eta}(z_1)\delta\eta(z_2)}.\nonumber
\label{eq:VERTEX3}
\end{eqnarray}
Using Eq.~(\ref{eq:currentvertexJmu}) and Eq.~(\ref{eq:3pcfomega}),
we obtain a relation
\begin{eqnarray}
&& \int dz_3 dz_4 G(z_1-z_3)\Gamma_{\gamma^{\mu}}(z,z_3,z_4)
G(z_4-z_2)\nonumber \\
&=& \int dz_3 dz_4 dz_5g_{\omega}^{-1} \left[(\partial^{2} +
m_{\omega}^{2})g^{\mu\nu}-\partial^{\mu}\partial^{\nu}\right]
\nonumber \\
&& \times F^{\rho}_{\nu}(z-z_5) G(z_1-z_3) \Gamma_{\rho}
(z_5,z_3,z_4)G(z_4-z_2).\nonumber \\
\label{eq:VERTEX4}
\end{eqnarray}
Under the condition of translational invariance, the current vertex
function $\Gamma_{\gamma^{\mu}}$ is Fourier transformed as
\begin{eqnarray}
\Gamma_{\gamma^{\mu}}(z_1,z,z_2) &\equiv&
\Gamma_{\gamma^{\mu}}(z_1-z, z-z_2) \nonumber \\
&=& \int\frac{d^4q d^4k}{(2\pi)^{8}}e^{-i(k+q)(z_1-z)} \nonumber \\
&& \times e^{-ik(z-z_2)}\Gamma_{\gamma^{\mu}}(q,k).\nonumber
\label{eq:VERTEX5}
\end{eqnarray}
Fourier transformation turns the relation (\ref{eq:VERTEX4}) into
\begin{eqnarray}
\Gamma_{\gamma^{\mu}}(q,k) = g_{\omega}^{-1}F^{-1}_{0\mu\nu}(q)
F^{~\rho}_{\nu}(q)\Gamma_{\rho}(q,k).\label{eq:VERTEX5}
\end{eqnarray}
It is more convenient to re-write this relation as
\begin{eqnarray}
F^{\mu\nu}(q)\Gamma_{\nu}(q,k) = g_{\omega}
F^{\mu\nu}_{0}(q)\Gamma_{\gamma_{\nu}}(q,k).
\label{eq:fmunuidentity}
\end{eqnarray}
Based on the invariance of $Z$ under arbitrary infinitesimal
variations of $\sigma$, $\rho^\mu_3$, and $\delta_3$ fields, one
could derive the following three exact identities
\begin{eqnarray}
D(q)\Upsilon(q,k) &=& -g_{\sigma} D_0(q)\Upsilon_{I}(q,k),
\label{eq:upsilonidentity} \\
V^{\mu\nu}(q)\Theta_{\nu}(q,k) &=& \frac{g_{\rho}}{2}V^{\mu\nu}_0(q)
\Theta_{\tau_3\gamma_{\nu}}(q,k),\label{eq:thetaidentity}
\\
C(q)\Omega(q,k) &=& -g_{\delta}C_{0}(q)\Omega_{\tau_3}(q,k).
\label{eq:omegaidentity}
\end{eqnarray}

\begin{figure}[htbp]
\centering
\includegraphics[width=3.36in]{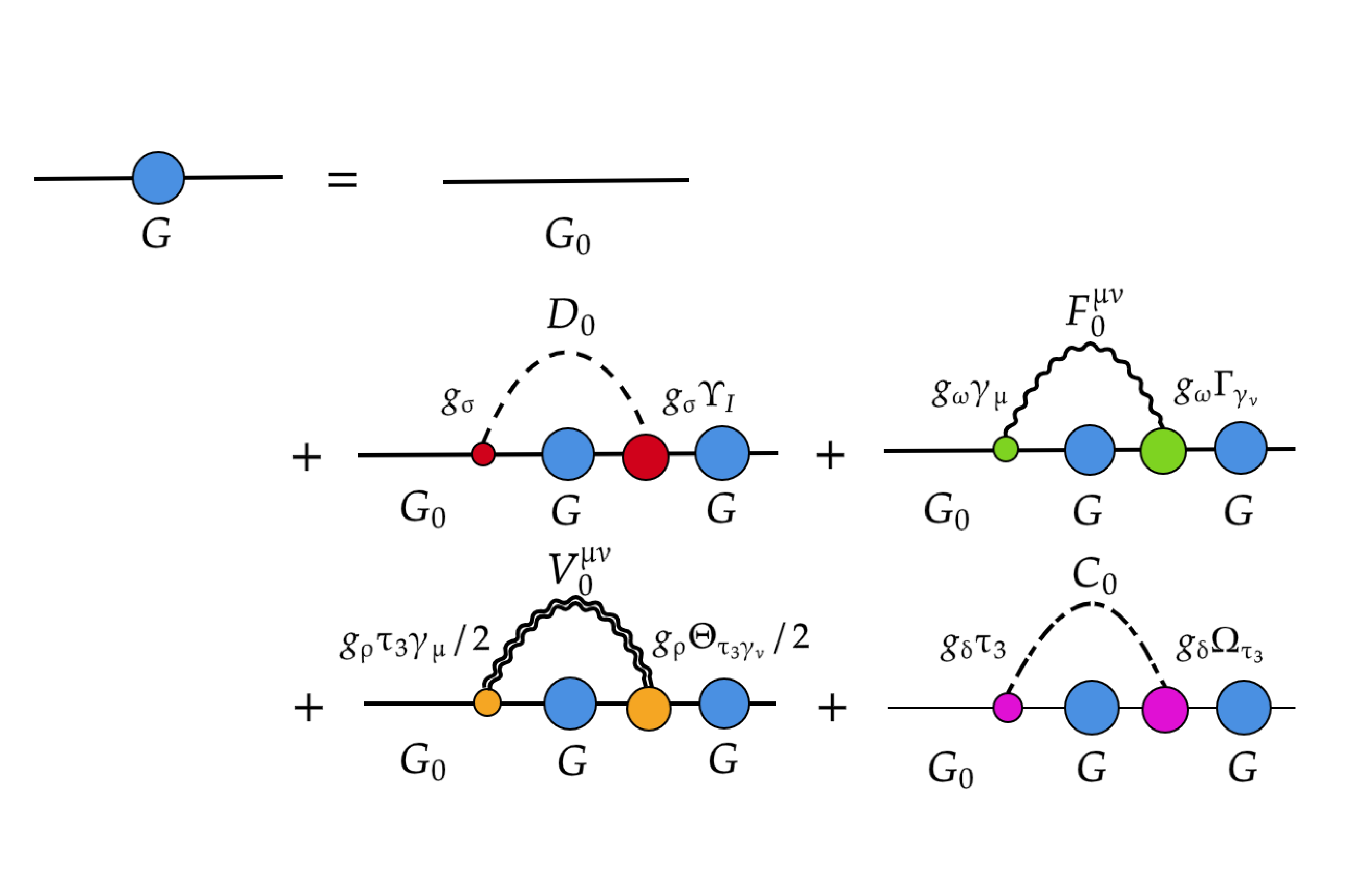}
\caption{A diagrammatic illustration of the exact DS equation of
$G(k)$ given by Eq.~(\ref{eq:dsegkfinal}).}
\label{fig:exactdsecurrent}
\end{figure}

Inserting the identities given by
Eqs.~(\ref{eq:fmunuidentity}-\ref{eq:omegaidentity}) into the DS
equation (\ref{eq:originaldseg}), we get
\begin{eqnarray}
G^{-1}(k) &=& G_0^{-1}(k) - ig^{2}_{\sigma}\int
\frac{d^4q}{(2\pi)^4} G(k+q)D_0(q)\Upsilon_{I}(q,k) \nonumber \\
&& -ig^{2}_{\omega}\int
\frac{d^4q}{(2\pi)^4}\gamma_{\mu}G(k+q)F^{\mu\nu}_{0}(q)
\Gamma_{\gamma_{\nu}}(q,k) \nonumber \\
&& -i\frac{g^{2}_{\rho}}{4}\int \frac{d^4q}{(2\pi)^4}\tau_{3}
\gamma_{\mu} G(k+q)V^{\mu\nu}_{0}(q)\Theta_{\tau_{3}
\gamma_{\nu}}(q,k)\nonumber \\
&& -ig^{2}_{\delta}\int\frac{d^4q}{(2\pi)^4}\tau_{3}
G(k+q)C_{0}(q)\Omega_{\tau_3}(q,k). \label{eq:dsegkfinal}
\end{eqnarray}
The Feynman diagram of this DS equation is shown in
Fig.~\ref{fig:exactdsecurrent}. It should be emphasized that this DS
equation is still exact, since it is derived from a number of
identities. But this equation cannot be directly solved because the
current vertex functions $\Upsilon_{I}(q,k)$,
$\Gamma_{\gamma_{\nu}}(q,k)$, $\Theta_{\tau_3\gamma_{\nu}}(q,k)$,
and $\Omega_{\tau_3}(q,k)$ are not known. These functions have a
complicated structure and cannot be expressed in terms of nucleon
and meson propagators.

To tackle the above equation, we now introduce a crucial
approximation. Let us focus on the expression
$g^{2}_{\sigma}\Upsilon_{I}(q,k)$. Before including vertex
corrections, the parameter $g_{\sigma}$ appearing in the Lagrangian
density takes a certain bare value. It would be renormalized by
vertex corrections to become $\tilde{g}^{2}_{\sigma}(q,k) =
g^{2}_{\sigma}\Upsilon_{I}(q,k)$. In principle, $\tilde{g}_{\sigma}$
depend on energy and momentum. For calculational simplicity, we
require $\tilde{g}_{\sigma}$ to be a constant, under the assumption
that it represents the mean value of $\tilde{g}^{2}_{\sigma}(q,k)$
obtained by averaging over all the allowed energies and momenta.
This allows us to make the replacement
\begin{eqnarray}
g^{2}_{\sigma}\Upsilon_{I}(q,k) \rightarrow
\tilde{g}^{2}_{\sigma}I_{4\times4}. \label{eq:rebaresigma}
\end{eqnarray}
The same manipulation can be applied to the other three coupling
parameters, namely
\begin{eqnarray}
g^{2}_{\omega}\Gamma_{\gamma_{\nu}}(q,k) &\rightarrow&
\tilde{g}^{2}_{\omega} \gamma_{\nu}, \label{eq:rebareomega}\\
g^{2}_{\rho}\Theta_{\tau_3\gamma_{\nu}}(q,k) &\rightarrow&
\tilde{g}^{2}_{\rho} \tau_3\gamma_{\nu},
\label{eq:rebarerho}\\
g^{2}_{\delta}\Omega_{\tau_3}(q,k) &\rightarrow&
\tilde{g}^{2}_{\delta}\tau_3.\label{eq:rebaredelta}
\end{eqnarray}
The renormalized coupling parameters $\tilde{g}_{\sigma}$,
$\tilde{g}_{\omega}$, $\tilde{g}_{\rho}$, and $\tilde{g}_{\delta}$
partially take into account the contributions of vertex corrections.
Their realistic values are determined by comparing the theoretical
results of $\rho^{}_{\mathrm{B}0}$, $E_{\mathrm{b}}$,
$m^{\ast}_{\mathrm{N}}$, $K$, and $E_{\mathrm{s}}$ calculated at the
saturation density to experimental data. The effects of vertex
corrections are inherently present in the experimental values of
observable quantities, and thus are naturally integrated into the
simulated values of $\tilde{g}_{\sigma}$, $\tilde{g}_{\omega}$,
$\tilde{g}_{\rho}$, and $\tilde{g}_{\delta}$. Purely for notational
simplicity, we still use the symbols $g_{\sigma}$, $g_{\omega}$,
$g_{\rho}$, and $g_{\delta}$ to denote the renormalized coupling
parameters. Formally, it is only necessary to make the following
replacement:
\begin{eqnarray}
\Upsilon_{I}(q,k) &\rightarrow& I_{4\times4}, \label{eq:baresigma}\\
\Gamma_{\gamma_{\nu}}(q,k) &\rightarrow& \gamma_{\nu},
\label{eq:bareomega}\\
\Theta_{\tau_3\gamma_{\nu}}(q,k) &\rightarrow& \tau_3\gamma_{\nu},
\label{eq:barerho}\\
\Omega_{\tau_3}(q,k) &\rightarrow& \tau_3.\label{eq:baredelta}
\end{eqnarray}
Then the DS equation (\ref{eq:dsegkfinal}) is simplified to
Eq.~(\ref{eq:dselwmodel}) presented in Sec.~\ref{sec:dse}.

%\begin{eqnarray}
%G^{-1}(k) &=& G_0^{-1}(k)-ig^{2}_{\sigma}\int
%\frac{d^4q}{(2\pi)^4} G(k+q)D_0(q)\nonumber \\
%&& -ig^{2}_{\omega}\int \frac{d^4q}{(2\pi)^4}\gamma_{\mu}
%G(k+q)F^{\mu\nu}_{0}(q)\gamma_{\nu} \nonumber \\
%&& -i\frac{g^{2}_{\rho}}{4}\int \frac{d^4q}{(2\pi)^4}\tau_3
%\gamma_{\mu}G(k+q)V^{\mu\nu}_{0}(q)\tau_{3}\gamma_{\nu}
%\nonumber \\
%&& -ig^{2}_{\delta}\int \frac{d^4q}{(2\pi)^4}\tau_3
%G(k+q)C_{0}(q)\tau_{3}, \label{eq:finalme}
%\end{eqnarray}

%If one is willing to treat the current vertex functions more
%carefully, it is necessary to consider the energy-momentum
%dependence of $g_{\sigma}$, $g_{\omega}$, $g_{\rho}$, and
%$g_{\delta}$. This is apparently very difficult given the intricate
%structure of current vertex functions. For a neutron near the Fermi
%surface, the different $|\mathbf{k}|- k^{~}_{\mathrm{F}}$ is a small
%quantity and can be used to expand the current vertex functions as a
%power series. For instance, $\Upsilon_{I}(q,k)$ can be expanded as
%\begin{eqnarray}
%\Upsilon_{I} &\approx& I_{4\times4} + b\left(|\mathbf{k}|-
%k^{~}_{\mathrm{F}}\right)I_{4\times4} + d\left(|\mathbf{k}|-
%k^{~}_{\mathrm{F}}\right)^{2}I_{4\times4} \nonumber \\
%&& +O(0^{3}). \label{eq:correbaresigma}
%\end{eqnarray}
%The coefficients $b$ and $d$ could be determined by fitting
%experimental data of nuclear quantities. Other current vertex
%functions $\Gamma_{\gamma_{\nu}}(q,k)$,
%$\Theta_{\tau_3\gamma_{\nu}}(q,k)$, and $\Omega_{\tau_3}(q,k)$ can
%be treated in an analogous way.


\begin{thebibliography}{99}
%\setlength{\itemsep}{0.5em}

\bibitem{Glendenningbook}
N. K. Glendenning, \emph{Compact Stars} (Springer, Berlin, 2000).

\bibitem{Lattimer04}
J. M. Lattimer and M. Prakash, The physics of neutron stars, Science
{\bf 304}, 536 (2004).

\bibitem{Lattimer16}
J. M. Lattimer and M. Prakash, The equation of state of hot, dense
matter and neutron stars, Phys. Rep. {\bf 621} 127 (2016).

\bibitem{Burgio21}
G. F. Burgio, H. J. Schulze, I. Vida$\tilde{n}$a, and J.-B. Wei,
Neutron stars and the nuclear equation of state, Prog. Part. Nucl.
Phys. {\bf 120}, 103879 (2021).

\bibitem{Walecka74}
J. D. Walecka, A theory of highly condensed matter, Ann. Phys.
(N.Y.) {\bf 83}, 491 (1974).

\bibitem{Boguta77}
J. Boguta and A. R. Bodmer, Relativistic calculation of nuclear
matter and the nuclear surface, Nucl. Phys. A {\bf 292}, 413 (1977).

\bibitem{Serot79}
B. D. Serot, A relativistic nuclear field theory with $\pi$ and
$\rho$ mesons, Phys. Lett. B {\bf 86}, 146 (1979).

\bibitem{Kubis97}
S. Kubis and M. Kutschera, Nuclear matter in relativistic mean field
theory with isovector scalar meson, Phys. Lett. B {\bf 191}, 399
(1997).

\bibitem{Reinhard89}
P.-G. Reinhard, The relativistic mean-field description of nuclei
and nuclear dynamics, Rep. Prog. Phys. {\bf 52}, 439 (1989).

\bibitem{Dutra14}
M. Dutra \emph{et al.}, Relativistic mean-field hadronic models
under nuclear matter constraints, Phys. Rev. C {\bf 90}, 055203
(2014).

\bibitem{Dutra16}
M. Dutra \emph{et al.}, Stellar properties and nuclear matter
constraints, Phys. Rev. C {\bf 93}, 025806 (2016).

\bibitem{NL3}
G. A. Lalazissis, J. Konig, and P. Ring, New parametrization for the
Lagrangian density of relativistic mean field theory, Phys. Rev. C
{\bf 55}, 540 (1997).

\bibitem{GM1}
N. K. Glendenning and S. A. Moszkowski, Reconciliation of
neutron-star masses and binding of the $\Lambda$ in hypernuclei,
Phys. Rev. Lett {\bf 67}, 2414 (1991).

\bibitem{TM1}
Y. Sugahara and H. Toki, Relativistic mean-field theory for unstable
nuclei with non-linear $\sigma$ and $\omega$ terms, Nucl. Phys. A
{\bf 579}, 557 (1994).

\bibitem{NL3omegarho}
C. J. Horowitz and J. Piekarewicz, Neutron star structure and the
neutron radius of ${}^{208}$Pb, Phys. Rev. Lett. {\bf 86}, 5647
(2001).

\bibitem{FSUGold}
B. G. Todd-Rutel and J. Piekarewicz, Neutron-rich nuclei and neutron
stars: A new accurately calibrated interaction for the study of
neutron-rich matter, Phys. Rev. Lett. {\bf 95}, 122501 (2005).

\bibitem{BigApple}
F. J. Fattoyev, C. J. Horowitz, J. Piekarewicz, and B. Reed,
GW190814: Impact of a $2.6$ solar mass neutron star on the nucleonic
equations of state, Phys. Rev. C {\bf 102}, 065805 (2020).

\bibitem{TW99}
S. Typel and H. H. Wolter, Relativistic mean field calculations with
density-dependent meson-nucleon coupling, Nucl. Phys. A {\bf 656 },
331 (1999).

\bibitem{DDME2}
G. A. Lalazissis, T. Nik$\check{s}$i$\acute{c}$, D. Vretenar, and P.
Ring, New relativistic mean-field interaction with density-dependent
meson-nucleon couplings, Phys. Rev. C {\bf 71}, 024312 (2005).

\bibitem{DD2}
S. Typel, G. R\"{o}pke, T. Kl\"{a}hn, D. Blaschke, and H. H. Wolter,
Composition and thermodynamics of nuclear matter with light
clusters, Phys. Rev. C {\bf 81}, 015803 (2010).

\bibitem{DDVT}
S. Typel and D. A. Terrero, Parametrisations of relativistic energy
density functionals with tensor couplings, Eur. Phys. J. A {\bf 56},
160 (2020).

\bibitem{DDMEX}
A. Taninah, S. E. Agbemava, A. V. Afanasjev, and P. Ring, Parametric
correlations in energy density functionals, Phys. Lett. B {\bf 800},
135065 (2020).

\bibitem{Waleckabook}
A. L. Fetter and J. D. Walecka, \emph{Quantum Theory of
Many-Particle Systems} (McGraw-Hill, New York, 2003).

\bibitem{Meissner}
E. Epelbaum, H. W. Hammer, and Ulf-G. Meissner, Modern theory of
nuclear forces, Rev. Mod. Phys. {\bf 81}, 1773 (2009).

\bibitem{Machleidt}
R. Machleidt and D. R. Entem, Chiral effective field theory and
nuclear forces, Phys. Rep. {\bf 503}, 1 (2011).

\bibitem{Carlson15}
J. Carlson \emph{et al.}, Quantum Monte Carlo methods for nuclear
physics, Rev. Mod. Phys. {\bf 87}, 1067 (2015).

\bibitem{Tews18}
I. Tews, J. Carlson, S. Gandolfi, and S. Reddy, Constraining the
speed of sound inside neutron stars with chiral effective field
theory interactions and observations, Astrophys. J. {\bf 860}, 149
(2018).

\bibitem{Huth22}
S. Huth, P. T. H. Pang, I. Tews, T. Dietrich, A. Le F$\grave{e}$vre,
A. Schwenk, W. Trautmann, K. Agarwal, M. Bulla, M. W. Coughlin, and
C. Van Den Broeck, Constraining neutron-star matter with microscopic
and macroscopic collisions, Nature (London) {\bf 606}, 276 (2022).

\bibitem{Abbott17}
B. P. Abbott \emph{et al.}, GW170817: observation of gravitational
waves from a binary neutron star inspiral, Phys. Rev. Lett. {\bf
119}, 161101 (2017).

\bibitem{Abbott18}
B. P. Abbott \emph{et al.}, GW170817: measurements of neutron star
radii and equation of state, Phys. Rev. Lett. {\bf 121}, 161101
(2018).

\bibitem{Miller19}
M. C. Miller, F. K. Lamb, A. J. Dittmann, S. Bogdanov, Z.
Arzoumanian, K. C. Gendreau, S. Guillot, A. K. Harding, W. C. G. Ho,
J. M. Lattimer, R. M. Ludlam, S. Mahmoodifar, S. M. Morsink, P. S.
Ray, T. E. Strohmayer, K. S. Wood, T. Enoto, R. Foster, T. Okajima,
G. Prigozhin, and Y. Soong, PSR J0030+0451 mass and radius from
NICER data and implications for the properties of neutron star
matter, Astrophys. J. Lett. {\bf 887}, L24 (2019).

\bibitem{Riley19}
T. E. Riley, A. L. Watts, S. Bogdanov, P. S. Ray, R. M. Ludlam, S.
Guillot, Z. Arzoumanian, C. L. Baker, A. V. Bilous, D. Chakrabarty,
K. C. Gendreau, A. K. Harding, W. C. G. Ho, J. M. Lattimer, S. M.
Morsink, and T. E. Strohmayer, A NICER view of PSR J0030+0451:
Millisecond pulsar parameter estimation, Astrophys. J. Lett.
{\bf887}, L21 (2019).

\bibitem{Miller21}
M. C. Miller, F. K. Lamb, A. J. Dittmann, S. Bogdanov, Z.
Arzoumanian, K. C. Gendreau, S. Guillot, W. C. G. Ho, J. M.
Lattimer, M. Loewenstein, S. M. Morsink, P. S. Ray, M. T. Wolff, C.
L. Baker, T. Cazeau, S. Manthripragada, C. B. Markwardt, T. Okajima,
S. Pollard, I. Cognard \emph{et al.}, The radius of PSR J0740+6620
from NICER and XMM-Newton data, Astrophys. J. Lett. {\bf 918}, L28
(2021).

\bibitem{Riley21}
T. E. Riley, A. L. Watts, P. S. Ray, S. Bogdanov, S. Guillot, S. M.
Morsink, A. V. Bilous, Z. Arzoumanian, D. Choudhury, J. S. Deneva,
K. C. Gendreau, A. K. Harding, W. C. G. Ho, J. M. Lattimer, M.
Loewenstein, R. M. Ludlam, C. B. Markwardt, T. Okajima, C.
Prescod-Weinstein, R. A. Remillard \emph{et al.}, A NICER view of
the massive pulsar PSR J0740+6620 informed by radio timing and
XMM-Newton spectroscopy, Astrophys. J. Lett. {\bf 918}, L27 (2021).

\bibitem{Demorest10}
P. Demorest , T. Pennucci, S. M. Ransom, M. S. E. Roberts, and J. W.
T. Hessels, A two-solar-mass neutron star measured using Shapiro
delay, Nature (London) {\bf 467}, 1081 (2010).

\bibitem{Antoniadis13}
J. Antoniadis , P. C. C. Freire, N. Wex, T. M. Tauris, R. S. Lynch,
M. H. van Kerkwijk, M. Kramer, C. Bassa, V. S. Dhillon, T. Driebe,
J. W. T. Hessels, V. M. Kaspi, V. I. Kondratiev, N. Langer, T. R.
Marsh, M. A. Mclaughlin, T. T. Pennucci, S. M. Ransom, I. H. Stairs,
J. van Leeuwen  \emph{et al.}, A massive pulsar in a compact
relativistic binary, Science {\bf 340}, 1233232 (2013).

\bibitem{Cromartie19}
H. T. Cromartie, E. Fonseca, S. M. Ransom, P. B. Demorest, Z.
Arzoumanian, H. Blumer, P. R. Brook, M. E. DeCesar, T. Dolch, J. A.
Ellis, R. D. Ferdman, E. C. Ferrara, N. Garver-Daniels, P. A.
Gentile, M. L. Jones, M. T. Lam, D. R. Lorimer, R. S. Lynch, M. A.
McLaughlin, C. Ng \emph{et al.}, Relativistic Shapiro delay
measurements of an extremely massive millisecond pulsar, Nat.
Astron. {\bf 4}, 72 (2019).

\bibitem{Romani22}
R. W. Romani, D. Kandel, A. V. Filippenko, T. G. Brink, and W.
Zheng, PSR J0952-0607: The fastest and heaviest known galactic
neutron star, Astrophys. J. Lett. {\bf 934}, L17 (2022).

\bibitem{Itzykson}
C. Itzykson and J.-B. Zuber, \emph{Quantum Field Theory}
(McGraw-Hill, New York, 1980).

\bibitem{Roberts00}
C. D. Roberts and S. M. Schmidt, Dyson-Schwinger equations: density,
temperature and continuum strong QCD, Prog. Part. Nucl. Phys. {\bf
45}, S1-S103 (2000).

\bibitem{Liu21}
G.-Z. Liu, Z.-K. Yang, X.-Y. Pan, and J.-R. Wang, Towards exact
solutions for the superconducting $T_c$ induced by electron-phonon
interaction, Phys. Rev. B {\bf 103}, 094501 (2021).

\bibitem{Pan21}
X.-Y. Pan, Z.-K. Yang, X. Li, and G.-Z. Liu, Nonperturbative
Dyson-Schwinger equation approach to strongly interacting Dirac
fermion systems, Phys. Rev. B {\bf 104}, 085141 (2021).

\bibitem{BCS}
J. Bardeen, L. N. Cooper, and J. R. Schrieffer, Theory of
Superconductivity, Phys. Rev. {\bf 108}, 1175 (1957).

\bibitem{MEtheory}
D. J. Scalapino, \emph{The electron-phonon interaction and
strong-coupling superconductivity}, in \emph{Superconductivity},
edited by R. D. Parks (Marcel Dekker, New York, 1969).

\bibitem{Luttinger}
J. M. Luttinger and J. C. Ward, Ground-state energy of a
many-fermion system.II, Phys. Rev. {\bf 118}, 1417 (1960).

\bibitem{MCMC}
D. Foreman-Mackey, D. W. Hogg, D. Lang, and J. Goodman, emcee: The
MCMC Hammer. Publ. Astron. Soc. Pac. {\bf 125}, 306 (2013).

%\bibitem{Baym71}
%G. Baym, C. Pethick, and P. Sutherland, The ground state of matter
%at high densities: equation of state and stellar models. Astrophys.
%J. {\bf 170}, 299 (1971).

\bibitem{Tolman34}
R. C. Tolman, Static solutions of Einstein's field equations for
spheres of fluid, Phys. Rev. {\bf 55}, 364 (1939).

\bibitem{Oppenheimer39}
J. R. Oppenheimer and G. M. Volkoff, On massive neutron cores, Phys.
Rev. {\bf 55}, 374 (1939).

\bibitem{Hinderer08}
T. Hinderer, Tidal Love numbers of neutron stars, Astrophys. J. {\bf
677}, 1216 (2008).%; {\bf 697}, 964(E) (2009).

\bibitem{Postnikov10}
S. Postnikov, M. Prakash, and J. M. Lattimer, Tidal Love numbers of
neutron and self-bound quark stars, Phys. Rev. D {\bf 82}, 024016
(2010).

\bibitem{Shen21}
X. Wu, S. Bao, H. Shen, and R. Xu, Effect of the symmetry energy on
the secondary component of GW190814 as a neutron star, Phys. Rev. C
{\bf 104}, 015802 (2021).

\bibitem{Annala18}
E. Annala, T. Gorda, A. Kurkela, and A. Vuorinen, Gravitational-wave
constraints on the neutron-star-matter equation of state, Phys. Rev.
Lett. {\bf 120}, 172703 (2018).

\bibitem{Abbott20}
R. Abbott, T. D. Abbott, S. Abraham, F. Acernese, K. Ackley, C.
Adams, R. X. Adhikari, V. B. Adya, C. Affeldt, M. Agathos, K.
Agatsuma, N. Aggarwal, O. D. Aguiar, A. Aich, L. Aiello, A. Ain, P.
Ajith, S. Akcay, G. Allen, A. Allocca \emph{et al.}, GW190814:
Gravitational waves from the coalescence of a $23$ solar mass black
hole with a $2.6$ solar mass compact object, Astrophys. J. Lett.
{\bf 896}, L44 (2020).

\bibitem{Prakash97}
M. Prakash, I. Bombaci, M. Prakash, P. J. Ellis, J. M. Lattimer, and
R. Knorren, Composition and structure of protoneutron stars, Phys.
Rep. {\bf 280}, 1 (1997).

\bibitem{Baiotti17}
L. Baiotti and L. Rezzolla, Binary neutron star mergers: a review of
Einstein's richest laboratory, Rep. Prog. Phys. {\bf 80}, 096901
(2017).

\bibitem{Li18}
J. J. Li, A. Sedrakian, and F. Weber, Competition between delta
isobars and hyperons and properties of compact stars, Phys. Lett. B
{\bf 783}, 234 (2018).

\bibitem{Sedrakian20}
A. Sedrakian, F. Weber, and J. J. Li, Confronting GW190814 with
hyperonization in dense matter and hypernuclear compact stars, Phys.
Rev. D {\bf 102}, 041301 (2020).

\bibitem{Sedrakian}
A. Sedrakian and J. W. Clark, Superfluidity in nuclear systems and
neutron stars, Eur. Phys. J. A {\bf 55}, 167 (2019).

\bibitem{Page}
D. Page, J. M. Lattimer, M. Prakash, and A. W. Steiner, Stellar
Superfluids, in \emph{Novel Superfluids}, edited by K.-H. Bennemann
and J. B. Ketterson (Oxford, 2014).

\bibitem{Annala20}
E. Annala, T. Gorda, A. Kurkela, J. N\"{a}ttil\"{a}, and A.
Vuorinen, A. Evidence for quark-matter cores in massive neutron
stars, Nat. Phys. {\bf 16}, 907 (2020).

\bibitem{Annala23}
E. Annala, T. Gorda, J. Hirvonen, O. Komoltsev, A. Kurkela, J.
N$\ddot{a}$ttil$\ddot{a}$, and A. Vuorinen, Strongly interacting
matter exhibits deconfined behavior in massive neutron stars, Nat.
Commun. {\bf 14}, 8451 (2023).

\end{thebibliography}
\end{document}